\newif\ifNotes\Notesfalse
\newif\ifAnon\Anonfalse
\newif\ifCamera\Cameratrue
\newif\ifConfidential\Confidentialfalse
\newif\ifCode\Codetrue

\newif\ifAmerican\Americanfalse

\ifCamera
    \Notesfalse
    \Anonfalse
    \Americanfalse
    \Codetrue
\fi

\newif\ifFlash\Flashtrue
\newif\ifThumb\Thumbfalse

\ifFlash
\else
    \Thumbtrue
\fi

\documentclass[letterpaper,twocolumn,10pt,table,url]{article}

\usepackage{usenix}

\usepackage[utf8]{inputenc}

\usepackage{graphicx}
\graphicspath{ {./images/} }
\usepackage[sort,numbers]{natbib}
\usepackage{tabularx}
\usepackage{tabu}
\usepackage{enumitem}
\usepackage[export]{adjustbox}
\usepackage{xcolor}
\usepackage{textgreek}
\usepackage{multirow}
\usepackage{bigstrut}
\usepackage{xspace}
\usepackage[caption=false]{subfig}
\usepackage{textcomp}
\usepackage{hyperref}
\hypersetup{
    colorlinks,
    linkcolor={red!50!black},
    citecolor={green!60!black},
    urlcolor={blue!80!black}}

\usepackage{amsmath,amssymb,amsfonts}
\usepackage{algorithmic}
\usepackage{graphicx}
\usepackage{textcomp}
\usepackage{xcolor}
\usepackage{booktabs}
\usepackage[capitalise,nameinlink,noabbrev]{cleveref}
\def\BibTeX{{\rm B\kern-.05em{\sc i\kern-.025em b}\kern-.08em
    T\kern-.1667em\lower.7ex\hbox{E}\kern-.125emX}}

\usepackage{fancyhdr}

\usepackage{xcolor,pifont}
\newcommand*\colourcheck[1]{
  \expandafter\newcommand\csname #1check\endcsname{\textcolor{#1}{\ding{52}}}
}
\colourcheck{blue}
\colourcheck{green}
\colourcheck{red}

\newcommand{\parhead}[1]{\vspace{1pt plus 2pt minus 1pt}\par\noindent\textbf{#1}\hspace{.4em plus .2em minus .2em}}

\ifNotes
    \newcommand{\colorcomment}[2]{\leavevmode\unskip\space{\color{#1}#2}\xspace}
\else
    \newcommand{\colorcomment}[2]{\leavevmode\unskip\relax}
\fi

\newcommand{\american}[2]{{\ifAmerican#1\else#2\fi\xspace}}

\newcommand{\doi}[1]{doi:~{\texttt{\href{https://doi.org/#1}{#1}}}}

\title{The Impostor Among US(B): Off-Path Injection Attacks on USB Communications} 

\makeatletter
\newcommand{\linebreakand}{
  \end{@IEEEauthorhalign}
  \hfill\mbox{}\par
  \mbox{}\hfill\begin{@IEEEauthorhalign}
}
\makeatother

\newcommand{\email}[1]{{\rm\textsf{\href{mailto:#1}{#1}}}}
\ifAnon
\author{}
\else

\author{
{\rm Robert Dumitru}\\
\rm \textit{The University of Adelaide} \&\\
\rm \textit{Defence Science and Technology Group}
\\\email{robert.dumitru@adelaide.edu.au}
\\\\
{\rm Andrew Wabnitz}\\
\rm \textit{Defence Science and Technology Group}
\\\email{andrew.wabnitz1@defence.gov.au}
\and
{\rm Daniel Genkin}\\
\rm \textit{Georgia Institute of Technology}
\\\email{genkin@gatech.edu}
\\\\\\
{\rm Yuval Yarom}\\
\rm \textit{The University of Adelaide}
\\\email{yval@cs.adelaide.edu.au}
} 

\fi

\begin{document}

\maketitle

\ifConfidential
\fancypagestyle{firstpage}
{
    \fancyhead[C]{Confidential draft -- Do not distribute}
}
\thispagestyle{firstpage} 
\else
\thispagestyle{empty}
\fi

\pagestyle{plain}

\begin{abstract}
USB is the most prevalent peripheral interface in modern computer systems and its inherent insecurities make it an appealing attack vector.
A well-known limitation of USB is that traffic is not encrypted. This allows \emph{on-path} adversaries to trivially perform man-in-the-middle attacks.
\emph{Off-path} attacks that compromise the confidentiality of communications have also been shown to be possible.
However, so far no off-path attacks that breach USB communications integrity have been demonstrated. 

In this work we show that the integrity of USB communications is not guaranteed even against off-path attackers.
Specifically, we design and build malicious devices that, even when placed outside of the path between a victim device and the host, can inject data to that path.
Using our developed injectors we can falsify the provenance of data input as interpreted by a host computer system.
By injecting on behalf of trusted victim devices we can circumvent any software-based \american{authorization}{authorisation} policy \american{defenses}{defences} that computer systems employ against common USB attacks.
We demonstrate two concrete attacks.
The first injects keystrokes allowing an attacker to execute commands.
The second demonstrates file-contents replacement including during system install from a USB disk.
We test the attacks on 29 USB 2.0 and USB 3.x hubs and find 14 of them to be vulnerable.

\end{abstract}

\section{Introduction}

The Universal Serial Bus (USB) has become the de-facto standard for computer-peripheral connection.
Since its original introduction in the late 90's, USB has replaced nearly all computer-peripheral connection standards.
Simplicity, ease of use, and enabling low-cost implementation have always been \american{prioritized}{prioritised} throughout the development of the standard, propelling its popularity over the past two decades.
Conversely, security has largely been overlooked throughout the development of USB.
With the USB Implementers Forum arguing that ``consumers should only grant trusted sources with access to their USB devices''~\cite{USBIFStatement}, USB's security model relies on restricting physical access, rather than on tried and tested techniques, such as permissions, encryption, and authentication.
In particular, operating systems typically immediately trust any USB device once connected, providing little feedback about the device's nature or capabilities.
Given the ubiquity of USB, it is important to understand and \american{characterize}{characterise} the attack surface and resultant threats presented by various usage and configuration scenarios. 
This \american{characterization}{characterisation} can further enhance secure usage of USB, mitigating a wide range of USB-based exploits.

\begin{figure}[t]
    \centering
    \includegraphics[width=\linewidth]{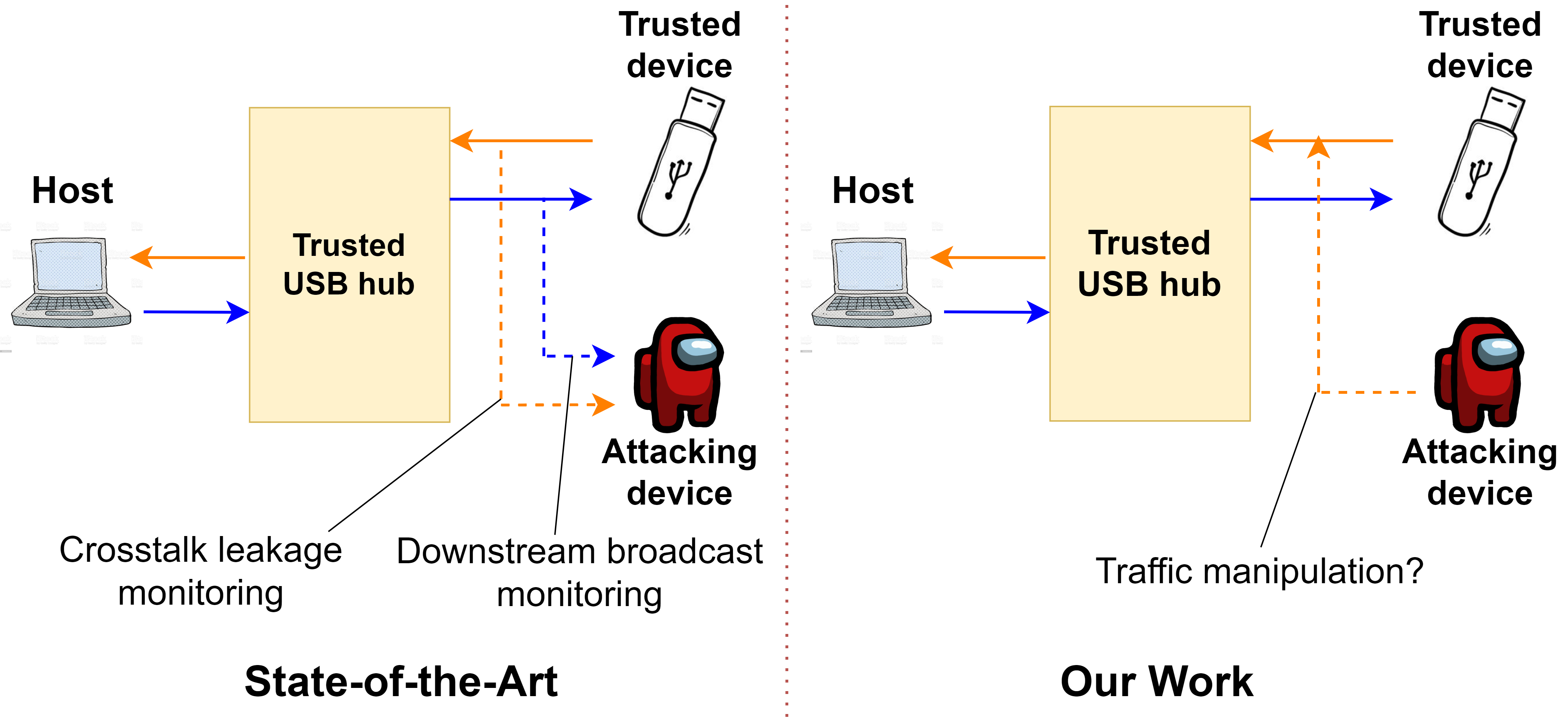}
    \vspace{-2em}
    \caption{Off-path attacks on USB communications: (left) off-path traffic snooping by monitoring broadcasts~\cite{NeugschwandtnerBK16} and using crosstalk leakage~\cite{SuGRY17}; (right) we show an off-path attack that  generates or manipulates the upstream traffic of other devices}
    \label{fig:on/off-path}
\end{figure}

With users often plugging untrusted USB devices into their computers~\cite{MATTHEW2016USERSREAL, JRJMasters}, numerous prior works have demonstrated attacks on the USB ecosystem via compromised devices.
Attacks range from \ifFlash flash \else thumb \fi drives compromising hosts by pretending to be keyboards~\cite{rubberducky}, to on-path entities such as hubs monitoring and manipulating USB traffic~\cite{badusb2mastersthesis, USBProxy}.
Beyond on-path attacks by devices with direct data access, USB is also vulnerable to off-path attacks, where an attacker's device is not located on a direct path between the victim and the USB host.
The left panel of \cref{fig:on/off-path} \american{summarizes}{summarises} the state-of-the-art for off-path attacks on USB.
In USB 1.x and USB 2.0, downstream traffic (host to devices) is broadcast on the bus making it observable for all devices on the bus traffic~\cite{NeugschwandtnerBK16}.
\citet{SuGRY17} demonstrate that off-path attacks on confidentiality of upstream traffic are also possible, allowing devices to observe USB traffic sent by devices on adjacent USB ports due to electrical crosstalk.

Given the feasibility of off-path attacks on the confidentiality of USB data~\cite{SuGRY17}, in this paper we set out to investigate the feasibility of off-path attacks on USB data integrity.
While this can be trivially achieved by on-path attackers (e.g., hubs)~\cite{badusb2mastersthesis}, we are not aware of any past demonstrations of an off-path attack that compromises the integrity of USB communication.
Thus, in this work, we ask the following questions:

\smallskip
\begin{centering}
  \emph{Does the USB protocol protect the integrity of upstream communication against a malicious off-path attacker? That is, can an untrusted malicious off-path device generate USB traffic that the host will attribute to a trusted device?}
\end{centering}

\begin{figure*}[!hbt]
    \centering
    \includegraphics[width=\linewidth]{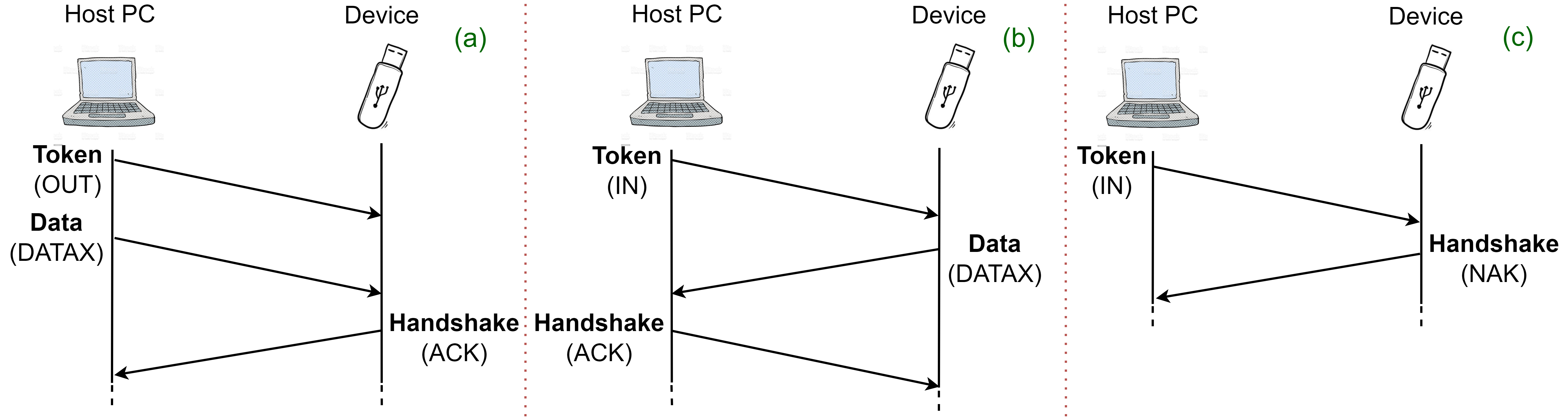}
\vspace{-2em}
    \caption{USB transaction examples: (a) host-to-device transaction, (b) device-to-host transaction, (c) device has nothing to send to host. The host first sends a token packet, identifying the target device within and the transfer direction. The host or the device then transmit data packets (with DATAX packet identifier). The handshake (or `status') packet terminates the transaction.}
    \label{fig:phases}
\end{figure*}

\subsection{Our Contribution}
In this work we show that USB does not protect the integrity of upstream communication even against off-path attackers.
More specifically, we describe an \emph{off-path USB injection} attack, 
which allows a malicious device located off the path between the victim device and host to send data which is accepted by the host as originating from the victim. See \cref{fig:on/off-path} (right). 

\parhead{Attack Mechanism.} Investigating the root cause of our attack, 
we find that when the host probes a device, the host and the hubs along the chain of connection fail to perform any verification that the response comes from the probed device.
This allows a malicious off-path device to respond when the host probes a different victim device, resulting in the transmitted data being accepted by the host as having originated from the victim.
We show how this mechanism can be used to send data to a host on behalf of devices that have been explicitly trusted by users through \american{authorization}{authorisation} policies, thereby bypassing common USB \american{defense}{defence} strategies.

\parhead{Attack Overview.}
In an attack scenario, an attacker-controlled device identifies as a benign USB device and performs the expected functionality for that device type.
In addition, the attacker's device also monitors all downstream USB traffic.
When it detects that the host probes the trusted victim device, it sends a malicious response that appears as if it were sent by the victim device.
At the same time, the victim device will also respond to the probe, creating a race condition between the malicious and the victim devices.
In the case that the malicious device wins the race, the host accepts the malicious response as if it were sent by the victim device.

\parhead{Attack Implementation.}
We implement a USB~1.x malicious device that identifies as a mouse but sends malicious keyboard input when the host probes a separate victim keyboard device.
We show that the attacker can consistently win the transmission race under vulnerable configurations, allowing a keystroke command injection attack.
We further build a USB~2.0 device that identifies as a serial communications device and monitors communications of a USB \ifFlash flash \else thumb \fi drive, replacing the contents of files that will reside on the host when they are transferred from the drive.
We show how this device can compromise a Linux installation.

\parhead{Bypassing Protection Policies.}
Where a host's USB stack has been instrumented with a defensive device \american{authorization}{authorisation} policy, our attack subverts this defense by falsifying provenance at the link layer.
This in turn allows us to exploit any trusted device interfaces or communication channels.
We show that our malicious devices bypass all \american{defenses}{defences} that restrict device function based on such policies.
We conclude with discussion on mitigation, recommendations for protecting existing systems and considerations for future system designs.

\parhead{Summary of Contributions.}
In summary, in this paper we make the following contributions:
\begin{itemize}[nosep,left=0pt]
  \item We identify a new attack on USB that allows off-path malicious devices to inject traffic to the communications between a trusted victim device and the host. (\cref{section:inj})
  \item We design and build two injection platforms, one capable of targeting USB 1.x devices and the other targeting USB~2.0 devices. (\cref{section:implementation})
  \item We investigate 29 USB~2.0 and USB~3.x hubs and find that 14 of them (48\%) are vulnerable to at least one form of attack. (\cref{sec:hubs}) 
  \item We demonstrate how our attack can inject keyboard payloads and mass-storage device data, specifically hijacking data transfer to the host. (\cref{section:keyboard,section:msd})
  \item We show that the attack bypasses prior \american{authorization}{authorisation} policy \american{defenses}{defences} aimed at preventing USB attacks such as device masquerading. (\cref{section:protections})
\end{itemize}

\subsection{Responsible Disclosure}\label{s:disclosure} Following the practice of responsible disclosure, we have shared our findings with the USB Implementers Forum (USB-IF), vendors of device \american{authorization}{authorisation} software, vendors of vulnerable hubs, and the manufacturers of their internal hub chips in order to disclose our findings.
At the time of writing there was no response from the USB-IF.
A report detailing the findings was sent to those who responded.
We note that the tested systems are all compliant with USB specifications, and the injection mechanism demonstrated in our work exploits a vulnerability in the protocol itself.
See \cref{table:disclosure} in the appendix for the responses.

\section{Background}
\label{section:backg}
First released in 1996, USB was designed to simplify the use of computer peripherals while replacing the plethora of then-common tailored interconnects with a single interface.
The primary simplification USB introduced was that it allowed automatic self-configuration of peripherals upon plugging in, referred to as `plug-and-play'.
USB 1.x~\cite{USB1.0, USB1.1} has two data transfer modes which we refer to collectively as \emph{classic-speed}: 1.5\,Mbps Low-Speed (LS) and 12\,Mbps Full-Speed (FS).
Human Interface Devices (HIDs) and other devices undemanding of bandwidth continue to be made as USB~1.x.

\parhead{USB 2.0.}
Later released in 2000, USB 2.0~\cite{USB2.0} is an extension of the USB specification which is capable of operating at 480\,Mbps in High-Speed (HS) mode.
Due to its increased data transfer speeds, USB~2.0 was able to meet the needs of high-bandwidth applications such as imaging, data acquisition systems and mass storage devices. 

\parhead{USB 3 and 4.}
USB 3.x~\cite{USB3.0, USB3.1, USB3.2}, initially released in 2008, is the latest major version of the protocol to have reached market maturity with speeds ranging from 5\,Gbps (SuperSpeed) to 
20\,Gbps (SuperSpeed Plus).
Backward compatibility with older device versions is built into USB 3.x host systems. 
Because of this and the extra cost of implementing a USB 3.x stack, devices only
support USB 3.x if they stand to benefit from the increased bandwidth. 
In particular, the interface for HID devices is defined over USB~1.x~\cite{USBHID}. 
Hence, these devices are unlikely to be implemented as USB 3.x.  The most recent version, USB~4~\cite{USB4}, released in 2019 operates at speeds of up to 40\,Gbps.
As of this writing there are a few devices marketed as this version.

\subsection{USB Communications}\label{sec:usb-comm}
USB systems are structured in a tree topology.
At the root is the host USB controller, which has an embedded \emph{root hub} that provides a tier of attachment points for peripheral devices.
Standard (non-root) hubs can be attached up to five in a chain to extend the number of USB ports, supporting up to 127 devices.
We simply refer to standard hubs as hubs.

USB communication is host-arbitrated and non-encrypted.
Downstream traffic originates from the host, which is broadcast in USB 1.x and 2.0, and unicast in later versions.
Upstream traffic is unicast to the host in all versions of the USB protocol.
The host manages the shared bus with poll-based Time-Division Multiplexing (TDM).

\parhead{Endpoints.}
Endpoints are essentially data sinks and sources through which USB devices communicate, usually implemented as hardware buffers on the device side and as pipes in host-side software.
All devices must support CONTROL endpoint 0, used for enumeration and status reporting.
Devices can support up to 15 additional IN and OUT endpoints.

\parhead{USB Transaction Protocol.}
All versions of USB use the same fundamental transaction protocol model.
Data communication through USB happens in transactions, which consist of up to three packet transmissions comprising \emph{token}, \emph{data} and \emph{handshake} phases.
The host always initiates transactions by sending a token downstream, containing 
the intended recipient's address, a packet identifier defining transaction type, and the endpoint number. 
According to the USB standard, devices must only process and respond to tokens addressed to them while ignoring others. 
Data is only transferred in one direction during a transaction, and the direction is specified in the token 
packet identifier,
with respect to the host.
For example, the host will send an OUT token to indicate it will transmit data to a device during the data phase, see \cref{fig:phases}(a).
Similarly, hosts use IN tokens to probe devices for input to be provided in a data phase
(c.f., \cref{fig:phases}(b)).
Otherwise, if devices have no data to send, they send a `NAK' handshake (c.f., \cref{fig:phases}(c)).

We note that no portion of the \emph{data} or \emph{handshake} phase packets identifies the source of the packet.
Instead the source is implicit in which device is addressed by the \emph{token} phase.

\parhead{USB 2.0 High Speed Extension of Transaction Protocol.}
As described above, during OUT transactions the host controls the bus for the majority of time until the handshake phase, at which point the device either ACKs or NAKs the received data.
This design can waste transmission time, for example in cases where the device is not prepared for data intake and thus NAKs the transaction. 
In order to mitigate this, USB 2.0 introduced an additional pre-transaction exchange before OUT communications at HS.
Here, the hosts first send a PING token to query devices as to whether it is ready to receive data.
The device can respond with a NYET (not yet) message or an ACK, in which case the host begins the OUT transaction.

\parhead{Backward Compatibility of USB 2.0 with USB 1.x.}
To reap the benefits of high bandwidth, in USB~2.0 systems all host--hub communication is delivered at HS (480Mbps), facilitated by Transaction Translator (TT) modules within USB~2.0 hubs.
TTs perform downstream speed mode translation and buffer transmissions from classic-speed devices for repeating upstream at HS.
Prior to transmitting LS or FS communications downstream at HS, a host will send a `SPLIT' packet indicating to the hub that it must translate the next incoming transmission to a 1.x speed mode before passing it on to the recipient device.
These SPLIT packets are sent prior to the token phase and only used in host--hub communication, making them not visible to end devices.
SPLIT packets include an information field specifying the hub to which the end recipient device is connected, resulting in only that hub translating the subsequent transmission.

\begin{figure}[h]
    \centering
    \includegraphics[width=0.55\linewidth]{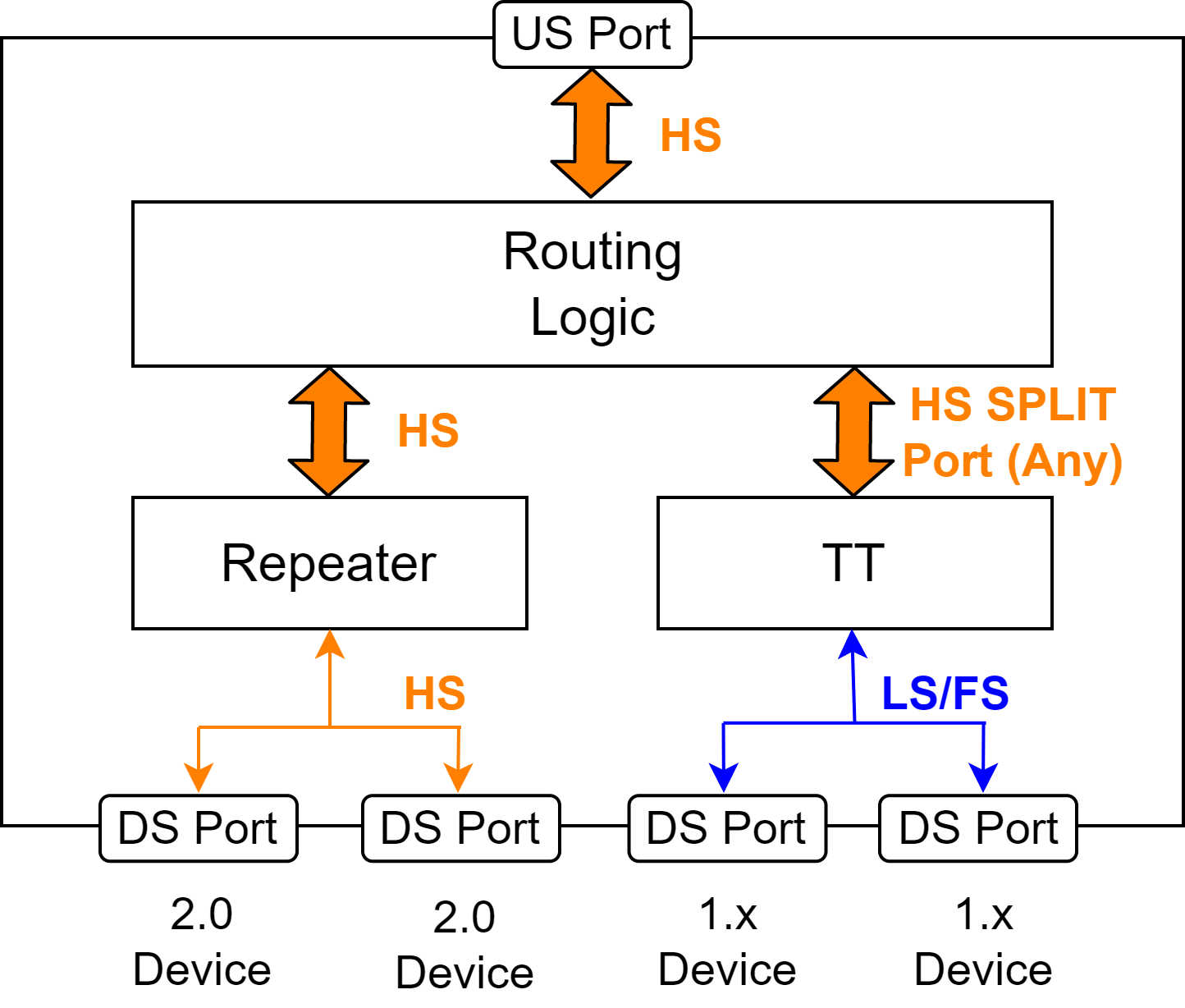}
    
    \bigskip
    
    \includegraphics[width=0.9\linewidth]{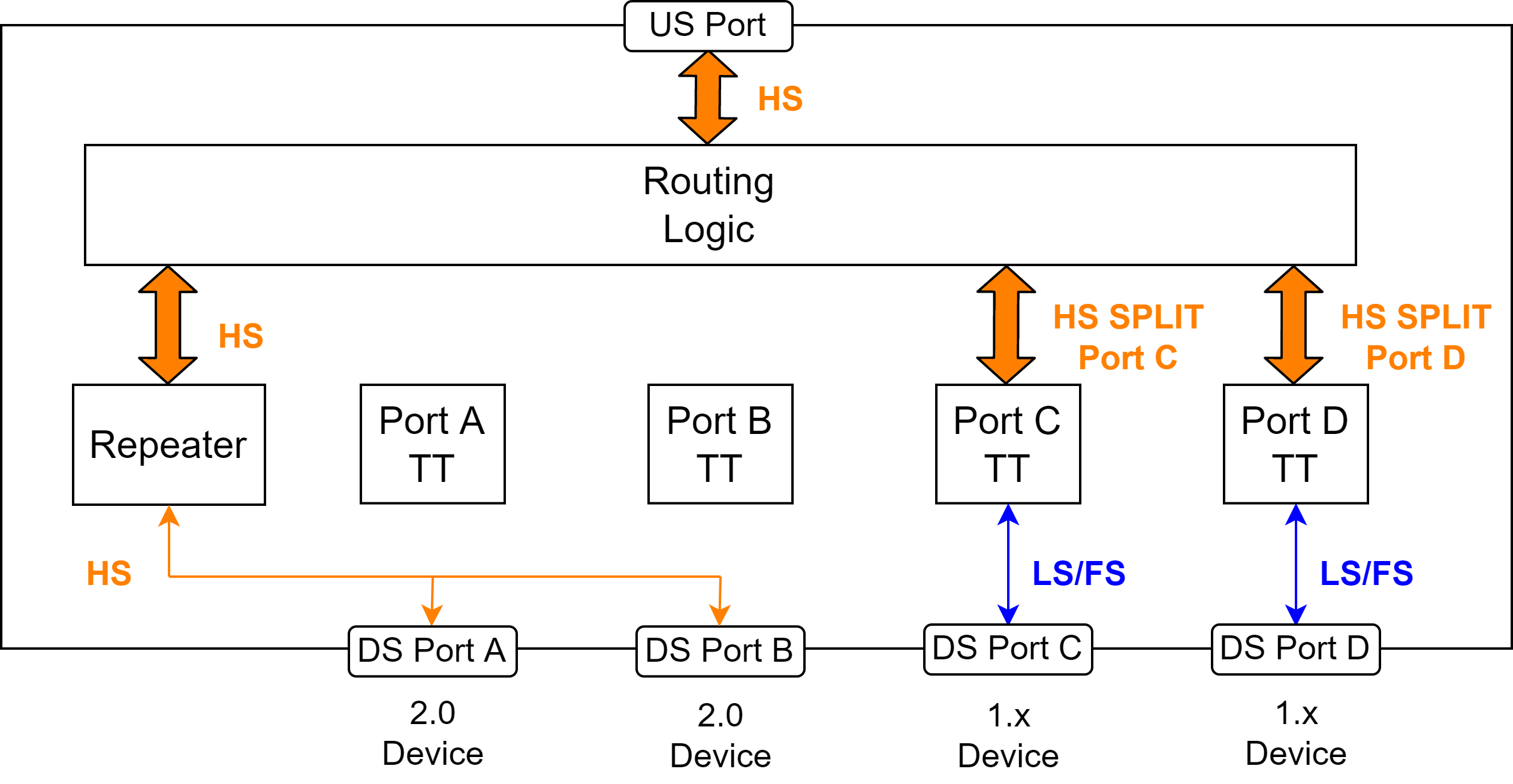}
\vspace{-1em}
    \caption{Traffic flows within single-TT (top) and multi-TT (bottom) hub with two (2.0) devices connected at HS and two (1.x) devices connected in classic-speed modes (LS or FS)}
    \label{fig:MultiTT}
\end{figure}

\parhead{USB Routing.}
The design for backward compatibility discussed above introduces a form of routing-to-hub for the translated classic-speed traffic.
Downstream classic-speed traffic is broadcast on the bus with a SPLIT header at HS, but then only translated at the target hub's TT for further broadcast at classic-speed.
On top of this, hubs can be designed as single-TT systems (see \cref{fig:MultiTT} top), where one TT handles all classic-speed traffic; or as multi-TT systems (see \cref{fig:MultiTT} bottom), where each downstream port has its own TT.
This also introduces routing to specific ports for classic-speed traffic through multi-TT hubs.

\parhead{Enumeration.}
USB enumeration is the process of identifying a recently plugged in device and establishing a connection between it and the host.
When a device is plugged in, the host will ask for its descriptor set containing self-reported (and not authenticated) information based on which the host can establish a connection.
The host will then set the necessary output power to the device, its speed mode, and loading appropriate drivers.
A newly connected device will use address 0 until the host assigns it a unique address early during enumeration. 

\begin{figure*}[h]
    \centering
    \includegraphics[width=0.7\linewidth]{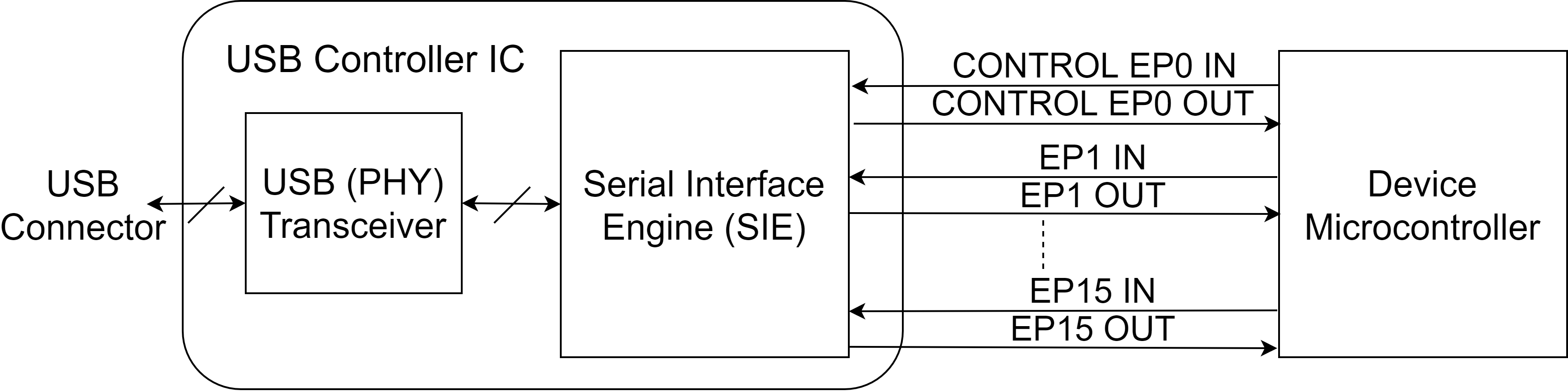}
    \vspace{-1em}
    \caption{Generic hardware architecture of a USB device. The USB Controller IC consists of a PHY and SIE. The PHY manages the translation between physical \american{analog}{analogue} signals and the logical data they represent, the SIE implements the link layer transaction protocol, including address checking, and the microcontroller runs the application that transfers data via USB to/from the host.}
    \label{fig:anatomy}
\end{figure*}

\subsection{Generic USB Device Architecture}\label{subsection:hardware}
\cref{fig:anatomy} shows the generic hardware architecture of a USB device.
The PHY (physical layer transceiver) manages physical bus activity, allowing for sending and receiving the serial signal on differential data lines.
The serial interface engine (SIE) 
module within the USB controller implements the transaction protocol, deals with time-critical operation and simplifies the microcontroller interface.
The SIE handles token address checking before demultiplexing and facilitating information exchanges for the various endpoints.

\subsection{Attacks on USB}\label{section:pastattacks}
The widespread adoption of USB among almost all PC, IoT, and embedded systems makes it an appealing avenue for exploitation.
Notably, USB's lack of any access control mechanisms leads to simple and effective attack vectors, while requiring costly and complicated countermeasures. 

We now review several categories of USB-based attacks and proposed protections, see~\cite{USBbasedattacks, USBsoftwareattacks, TianSKBBB18, exploringusbattacktaxo, USBSideChannels} and references therein for more complete descriptions.

\parhead{Obtaining Access.}
USB-based attacks rely on physical access to target systems and thus restriction of access to trusted devices is often used as justification for not securing USB systems.
More specifically, the USB Implementers Forum (USB-IF) released a \emph{Statement regarding USB security}~\cite{USBIFStatement}, which says: ``consumers should only grant trusted sources with access to their USB devices''.
However, typical USB-based attacks have minimal hardware requirements (size-wise) and malicious devices are often able to obscure their real \american{behavior}{behaviour} from unsuspecting users.
This results in users connecting devices from unknown origins to their systems, either inadvertently (e.g., supply chain compromise), naturally~\cite{MATTHEW2016USERSREAL}, or as a result of social engineering~\cite{JRJMasters}.

\parhead{Attacks on USB Confidentiality and Integrity.}
USB's lack of any encryption and authentication mechanisms allows adversaries to eavesdrop
on USB traffic, by both on-path and off-path entities. 
On-path or `in-line' entities are those on the chain of connection between host and the subject device.
Thus, attacks based on them require a stronger adversarial model than off-path attacks, as off-path entities do not require direct access to the targeted USB traffic. 

\parhead{On-Path Attacks.}
On-path entities can be passive devices such as protocol \american{analyzers}{analysers} acting as wiretaps.
Alternatively, they might also be active devices such as USB hubs which repeat communications passing through from one side to the other.
Hardware key loggers~\cite{keycarbon, keelog} are devices typically implemented as hubs with additional capability to record all keystroke traffic forwarded to a host, enabling capture of potentially sensitive information such as passwords.
BadUSB 2.0~\cite{badusb2mastersthesis} is a hub which can compromise USB communications by performing full man-in-the-middle (MITM) attacks, including eavesdropping, modifying, replaying, fabricating, and even exfiltrating data sent between hosts and devices.
Similarly, USBProxy~\cite{USBProxy} is an embedded system that uses USB On-The-Go~\cite{USBOTG}\footnote{Supplement to the USB specification that lets devices also assume the role of hosts, largely used in smartphones} controllers to act as a MITM.

\parhead{Off-Path Attacks.}
As all USB 1.x and 2.0 downstream traffic is broadcast on the bus, 
Neugschwandtner et al.~\cite{NeugschwandtnerBK16} use a protocol \american{analyzer}{analyser} connected as an off-path device to directly monitor downstream communications to all connected devices.
Such transmissions could include sensitive file contents in transit to a storage device, network adapter, or printer.

Similarly, upstream transmissions have been demonstrated to be observable by off-path devices~\cite{SuGRY17} due to a `crosstalk leakage' effect exhibited by the large majority of tested USB~2.0 hubs.
By monitoring these leaked signals from adjacent USB ports, off-path devices can eavesdrop on unicast upstream transmissions of other devices to the USB host.

\parhead{Electrical Attacks.}
Devices such as \emph{USB Killer}~\cite{USBkiller} can permanently incapacitate host computers and other attached devices by discharging high voltage direct current over the USB data lines, damaging all connected circuitry.
Robust electrical design protects against such attacks.
This involves opto- or galvanic isolation of the USB port circuitry from the connected host controller or hub.

\parhead{Device Masquerading.}
USB device controllers can also be programmed to emulate the operation of certain devices, \american{capitalizing}{capitalising} on the lack of device authentication and USB's ``plug-and-play'' nature. These so called \emph{masquerading attacks} use seemingly innocuous USB sticks with their firmware modified to emulate HID keyboards~\cite{rubberducky, driveby, USBcovertchannels, urfuked}.
Being able to feed arbitrary keystroke input enables adversaries to compromise computer systems in many different ways.
Payloads of keystroke sequences can also be pre-loaded on-chip~\cite{rubberducky}.
Alternatively, \emph{URFUKED}~\cite{urfuked} (the Universal RF USB Keyboard Emulation Device) incorporates RF communication for short range control over the keyboard emulator. Such attacks have become increasingly accessible with the advent of software-defined \american{customizable}{customisable} USB devices~\cite{rubberducky, goodfet}.

\subsection{\american{Defenses}{Defences}}
As outlined above, USB's design has mostly overlooked security aspects, with the USB-IF leaving the inclusion of security measures to the discretion of system implementers~\cite{USBIFStatement}.
This has resulted in a fragmented collection of bespoke \american{defenses}{defences}~\cite{TianSKBBB18}, most of which are only 
enacted at one given layer within the USB stack.
This is alarming, as \mbox{\citet{TianSKBBB18}} present many attacks which transcend multiple layers of the USB protocol stack, highlighting issues with non-holistic security solutions.

\parhead{Policies.}
Device \american{authorization}{authorisation} policies~\cite{usblockrp, 10.1145/3289100.3289121, USBGuard, windows8.1filter, 197251, ivanti, virtualbox, GoodUSB} are a widely adopted protection against USB-based threats. These offer users varying degrees of control over device function.
Protection policies vary from filtering communications to only allow certain devices, allowing certain interfaces within devices, and even only allowing certain interfaces to communicate with specific processes running on the host~\cite{10.1145/3289100.3289121}.
These policies also tend to fingerprint devices based on non-authenticated information that the device self-reports during enumeration.
A survey conducted in 2019 found that over 40\% of \american{organizations}{organisations} apply protection policies~\cite{apricorn}.

\parhead{Proxy Devices.}
Protective proxy devices~\cite{condom, highseclabs, USBSentry, USBArmadillo} can be placed between host and device in order to filter out \american{unauthorized}{unauthorised} traffic. As these devices also 
provide physical separation between the attacker and the system's data lines, proxy devices also offer incidental protection against bus sniffing attacks. However, to be effective, proxy devices must be implemented in conjunction with complete system retrofits that physically disable other ports~\cite{usblockrp, kensington, smartkeeper}.

\parhead{Cryptographic Protections.}
Cinch~\cite{197175} introduces a cryptographic overlay protocol to augment USB with (bidirectional) encryption and authentication. While this approach protects against off-path confidentiality breaches, it also introduces more than 80\% performance degradation due to  throughput reduction.
These types of \american{defenses}{defences} incur additional burden as they require instrumentation of host USB stacks.

\section{Threat Model}
In our injection threat model there are at least two USB devices connected to a common host via some USB topology of USB hubs. 
We assume that one of the devices is our malicious {injection platform}, which is under the attacker's control.
We further assume that the system contains a victim device, which is outside of the attacker's influence.
This is the device whose communications the attacker would like to impersonate.
Crucially, as per the tree topology of USB systems, the injection platform and the victim have logically separate, but physically shared, communication paths to the host.
The system might have additional USB devices attached.
These are considered bystanders, outside of the attacker's influence, and perform their function with unaffected USB communications.

\parhead{Other System Components.}
We assume that all of the system components, except for the injection platform, are trustworthy.
In particular, this includes the hubs on the path from the victim to the host as well as the host's operating system.
We also assume that there are no malicious or compromised on-path entities which might help the injection platform to impersonate the victim. 

\parhead{Hardened Hosts.}
While not typically implemented by default on computer systems, we allow the host to employ a device \american{authorization}{authorisation} policy in its USB software stack.
Such policies limit the types of devices that the system supports and allows, e.g.\ via an \american{authorization}{authorisation} list.
Alternatively or additionally, the policy may require some form of user approval when a new device is plugged in~\cite{GoodUSB}.
The policy may restrict the nature of the communication with the device, e.g.\ to filter out malicious communication or to ensure that traffic is consistent with the communication protocol of the \american{advertized}{advertised} device type. 

\parhead{Correct Implementation of \american{Authorization}{Authorisation} Tools.}
In case such an \american{authorization}{authorisation} policy is present, we assume that these tools are correctly implemented and deployed, fingerprinting devices by any means at its disposal in the software stack.
We also assume that such policy is correctly configured in that it accurately represents the user's intentions.
We assume that the user and the policy trust the trusted victim device and allow its communication.

\parhead{Policy Assumptions for Hardened Hosts.}
While the above assumes the correct functionality of 
\american{authorization}{authorisation} tools, we do not assume any specific USB policy with respect to the injection platform, provided that it is physically connected to the host.
The \american{authorization}{authorisation} policy may even completely disallow communication with the injection platform, only allowing the USB communication with specific hand-picked trusted devices. 

\section{Attack Overview}
\label{section:inj}

\begin{figure}[htbp]
    \centering
    \includegraphics[width=\linewidth]{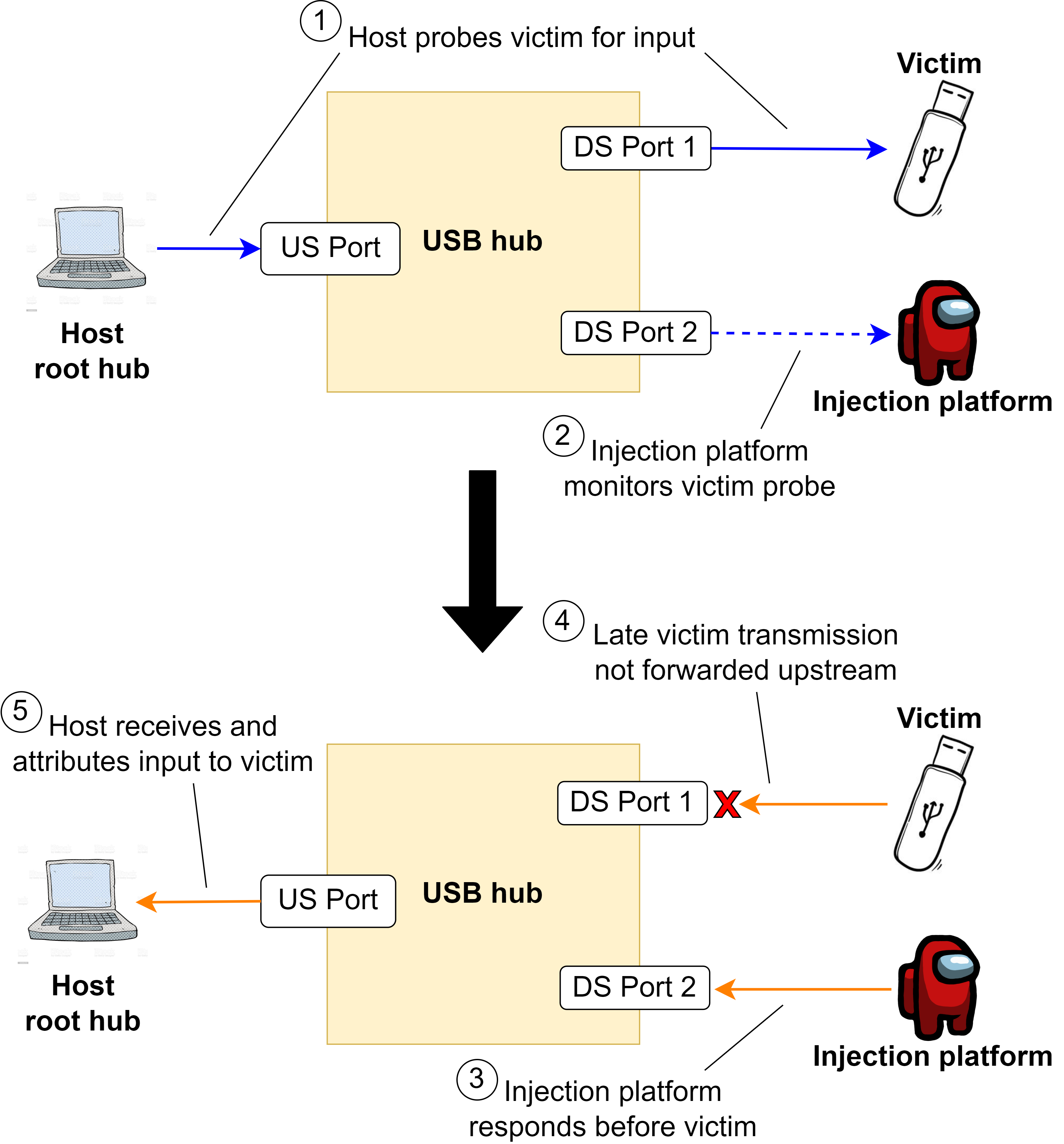}
\vspace{-2em}
    \caption{Injection arrangement and stages}
    \label{fig:Two step}
\end{figure}

\label{subsection:attackoverview}
Our injection platform displays two types of \american{behaviors}{behaviours}. 
It primarily functions as an innocuous USB device in its own right.
Additionally, it inconspicuously injects upstream communications data to the bus, aiming to impersonate the victim.
To that aim, we equip our injection platform with the ability to monitor the host's downstream communications for probes addressed to the victim, which trigger injections.

\parhead{Injection.}
\cref{fig:Two step} presents an overview of our injection attack.  First, the host broadcasts a probe requesting input from the victim \raisebox{.5pt}{\textcircled{\raisebox{-.9pt} {1}}}.
The injection platform observes the host's probe \raisebox{.5pt}{\textcircled{\raisebox{-.9pt} {2}}} and responds with an upstream data transmission \raisebox{.5pt}{\textcircled{\raisebox{-.9pt} {3}}} which matches the format of the expected victim response. 
Such \american{behavior}{behaviour} is in violation of the USB specification.
However, if the injection platform manages to respond before the victim device, the hub may accept the injected transmission and forward it upstream, while ignoring the victim's genuine response \raisebox{.5pt}{\textcircled{\raisebox{-.9pt} {4}}}.
Finally, we note that USB data and handshake responses do not carry address information.
Thus, when a response arrives at the host, the host cannot distinguish between sources based on the received data, rather it attributes a response to the most recently probed device.
Overall, in the case that the USB hub forwards the injected response upstream, the response from our injection platform is automatically attributed to the victim \raisebox{.5pt}{\textcircled{\raisebox{-.9pt} {5}}}.

\parhead{Bypassing Policies.}
Our attack exploits a vulnerability in the USB protocol brought about by a specification compliance assumption.
Using our injection technique we can circumvent the USB \american{authorization}{authorisation} policies on the host's hardware, as the host's USB controller cannot verify the source of the injected USB traffic. 
The crucial follow-on effect is that injection also bypasses software-based protection policies, as these implicitly trust the communications received from the host's USB controller to be correctly attributed.

\begin{figure}[h]
    \centering
    \includegraphics[width=0.90\linewidth]{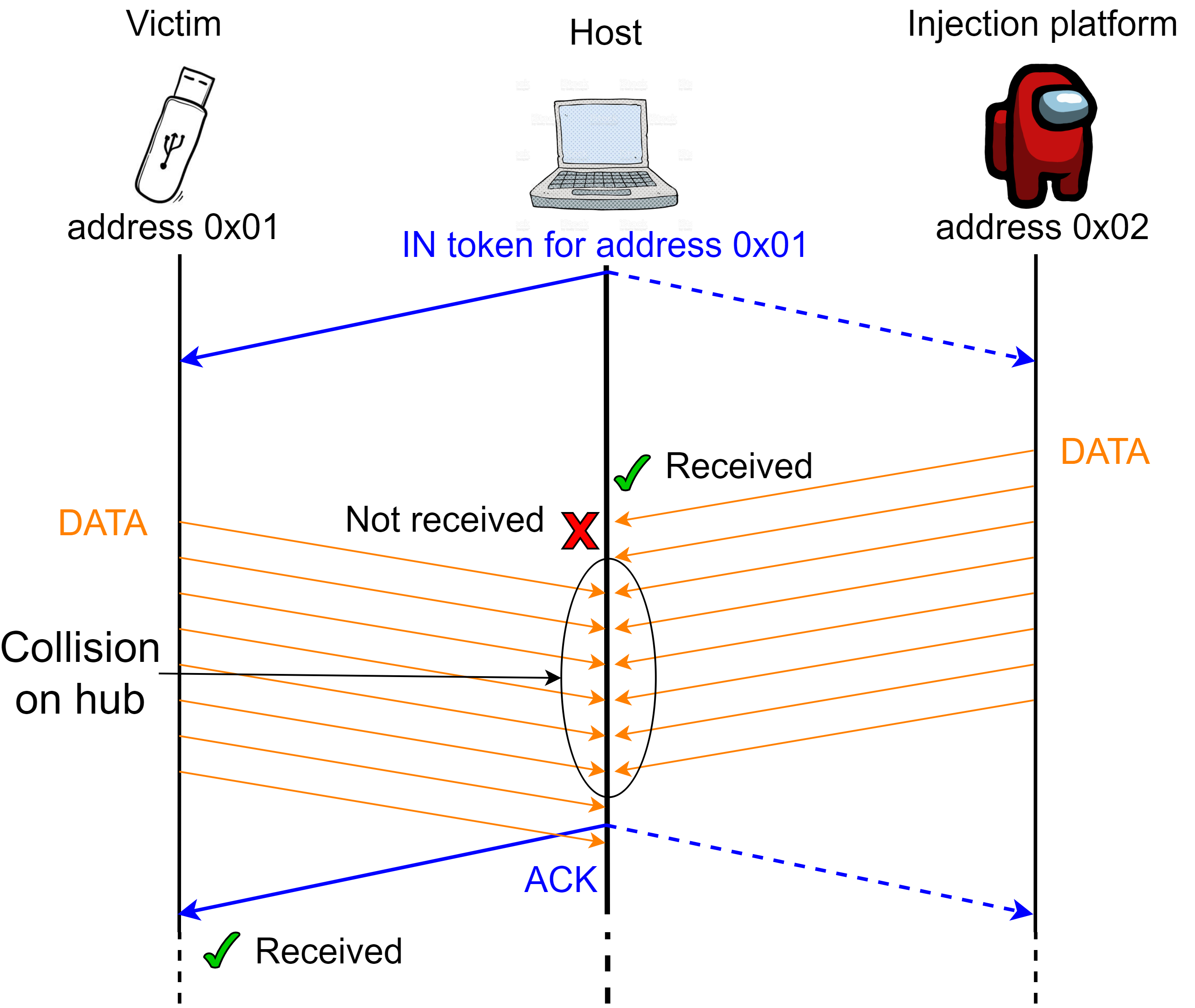}
    \caption{Sequence diagram of injection transmission and its collision with genuine victim transmission. 
        Dotted lines represent transmissions that are observable by the injection platform despite it not being the intended recipient.
        Although the sequence diagram shows multiple arrows for DATA phases, these are part of singular continual transmissions only shown this way to represent the long transmission period.}
    \label{fig:SequenceCollision}
\end{figure}

\parhead{Transmission Collisions on Hubs.}
Our attack exploits a race between the victim and the injection platform.
If the injection platform manages to send a response before the victim starts sending its response, the attacker clearly wins the race and injects their response.
However, injecting the whole response before the victim starts transmitting is quite unlikely, particularly when the attacker wishes to inject large amounts of data.
When the transmissions of the victim and the injection platform collide, the hub needs to handle the collision.
\cref{fig:SequenceCollision} depicts such a DATA -- DATA collision. 

\parhead{Collision Resolution.}
In the case of a collision, the USB specification~\cite{USB2.0} permits two \american{behaviors}{behaviours}: a hub can treat the later transmissions as errors, completely ignoring them.  Alternatively, the hub can detect the collision and send a `garble' error message upstream to the host.

Hubs that exhibit the former \american{behavior}{behaviour}, i.e.\ ignoring and discarding the later of incoming transmissions, are vulnerable to our injection attack.
With collision-detecting hubs, injection can still effect a Denial-of-Service (DoS) against victim devices, blocking them from providing input.

\section{Injection Platform Implementation}
\label{section:implementation}
Having outlined the general principles behind our USB injection attack, in this section we describe the implementation of our injection platforms.

\parhead{Triggering Injection.}
In USB 1.x and 2.0 systems, downstream communications are broadcast and can therefore be monitored directly by all devices in the USB topology, including off-path devices.
Our injection platforms seek specific patterns in downstream traffic, which upon detection prompt them to inject traffic upstream.
At a minimum, the final part of the expected pattern will consist of an IN token addressed to the victim device, which the host uses to probe it for input.
As detailed in~\cref{section:inj}, the host will attribute received data according to whichever device was most recently probed.

Many classes of device implement their own communication protocol on top of the USB protocol, typically using multiple transactions and multiple endpoints.
For example, hosts will request certain data from storage devices by issuing commands downstream using OUT endpoint transactions.
To effect these higher level communications, our injection platform must \american{recognize}{recognise} the relevant message sequence and subsequently trigger injections according to exchanges \american{contextualized}{contextualised} by the victim device class protocol, using packets crafted to match the corresponding format.

\parhead{Existing Device Implementations.}
As described in~\cref{section:backg}, the USB transaction protocol is typically implemented by a dedicated SIE hardware module within a device's USB controller.
Its function includes handling address checks of incoming tokens and the subsequent processing, i.e. when the token matches its device's address the SIE will write data to an OUT endpoint buffer or read data from an IN endpoint buffer.
An existing device implementation can be turned into an injection platform through modification of its SIE, specifically the SIE's token address check function.

\parhead{Target Device Type.}
To implement our injection platforms we modified hardware implementations of device SIEs within their USB controllers.
This involved modifying the RTL source of cores implementing USB devices.
Working at the hardware level offered high-fidelity control over timing which helped us ensure that our platforms win transmission races as required for the attack.
As a possible alternative solution, there are some general purpose microcontroller families~\cite{MCUs} with USB connectivity which implement the SIE in software/firmware and either support the USB interface directly (up to FS) or through an external PHY.
While modifying the firmware of such an implementation could be a viable means of achieving our objectives, we have pursued hardware-based solutions instead because of the greater control (due to less abstraction) they offer over the platform function.

\parhead{Hardware Setup.}
We have built two prototype injection platforms using existing implementations of USB device cores, which we deploy on FPGAs. 
The platforms are based on USB~1.x and~2.0 device implementations, one for each major version since the versions have different electrical interfaces and slightly different hardware \american{behaviors}{behaviours}.
While they are largely similar, some properties of the devices differ between the implementations.
We list these differences in \cref{table:injplatforms}.

\begin{table}[ht]
  \small
\centering
  \begin{tabular}{@{}lll@{}}
 &  \textbf{USB 1.x} &  \textbf{USB 2.0}\\
\midrule 
 Original core name & FPGA-USB-V2 & USB2SOFT USB 2.0\\
 & project~\cite{USB2FPGA} & device SIE~\cite{comblockcore} \\
 Speed modes &  LS (1.5Mbps) or &  HS (480Mbps)\\ 
 &  FS (12Mbps) &  \\ 
  Interface to USB & Direct to USB &  12-pin UTMI+ to\\
  & connector & PHY Waveshare\\
PHY &  Internal & USB3300 \\
Adapted core from & Open source & Licensed IP\\
RTL source language & VHDL & VHDL\\ 
FPGA target & Xilinx Artix-7 & Xilinx Kintex-7\\ 
Development board & Digilent Basys 3 & Digilent Genesys 2\\
\bottomrule
\end{tabular}
\vspace{-1em}
  \caption{Injection platform properties}
\label{table:injplatforms} 
\end{table}

\subsection{USB 1.x Injection Platform}\label{s:usb1.x}
\parhead{Core.}
The classic-speed (1.x) injector platform we have created is based on a modified (and debugged) core~\cite{USB2FPGA} written in VHDL
\ifCode
, which is freely available for distribution.\footnote{
\ifAnon 
  We will make the modified injection platform source RTL and bitstream available for artifact evaluation and will publish them as open source.
\else 
  The injection platform bitstream and source RTL code are available at: \url{https://github.com/0xADE1A1DE/USB-Injection}
\fi
}
\else
.
\fi
The core contains elements that perform the functions of all hardware modules within a device controller as in \cref{fig:anatomy} (PHY and SIE), however its structure is monolithic rather than partitioned into those distinct modules.
In conventional devices, communication across all endpoints would be handled by the device microcontroller.
This core instead defines functionality for device enumeration, performed by communication over control endpoint (0), directly in hardware.
The core also allows for communication over input/output (non-control) endpoints to be implemented directly in hardware.
The core interfaces directly with the USB differential data lines D+ and D- and can be set to operate as either a LS or FS device.

\parhead{Deployment.}
We instantiated the core on a Xilinx Artix-7 FPGA housed on a Digilent Basys 3 development board.
We directly connect the USB data lines from a spliced USB cable to 3.3V general-purpose I/O pins on the board.
The device is shown in \cref{fig:LSplat} in the appendix.

To configure the device to work at LS, we pull up the D- line to 3.3V across a 1.5k\textOmega~resistor, this signals the speed mode to the host and sets the associated differential signalling polarity.
To work at FS, the D+ line must be pulled up instead.

\subsection{USB 2.0 Injection Platform}\label{s:usb2.0}

\parhead{Core.}
The HS injection platform is based on an adapted licensed device core IP~\cite{comblockcore} written in VHDL.
This core instantiates a SIE and implements enumeration functionality, handling all CONTROL interactions over endpoint 0.
FIFO interfaces support IN and OUT endpoints.

The core connects to an external PHY across the low pin count interface -- UTMI+ (ULPI)~\cite{ULPI}.
This is the de facto standard interface for SIEs interacting with USB transceivers.
We use the interface with an 8-bit wide data bus and control signals which are clocked by the PHY with a 60MHz signal derived from transitions on the 480Mbps HS data lines.

\parhead{Deployment.}
We have ported the core to a Xilinx Kintex-7 FPGA housed on a Digilent Genesys 2 development board.
The board is connected through its general-purpose I/O pins, which are capable of switching at high frequency, to the UTMI+ pins of a Waveshare USB3300 PHY board.
\cref{fig:HSplat} in the appendix shows this device.

\subsection{SIE Modifications}
As outlined in \cref{sec:usb-comm}, Endpoint 1 is typically the main input (IN) endpoint used by devices and endpoint 0 is the CONTROL endpoint used for conveying setup information during enumeration.
Since we do not wish to interfere with our victim's enumeration, we configure our platforms to only inject endpoint 1 traffic.

As an example, in the USB 1.x device implementation RTL source we identify the following line which defines the device's address check \american{behavioral}{behavioural} logic on incoming tokens:

\medskip
\noindent \texttt{if (token\_ad /= new\_usb\_addr) or \\\indent(pid /= not token\_in(16 downto 13) then}
\medskip

\noindent where `\texttt{/=}' is the VHDL inequality operator.
When this \texttt{if} statement evaluates as true due to an address mismatch (first condition) or transmission error (second condition), the currently inspected token packet is no longer processed and the device waits for the next token.
We transform the device into an injection platform by modifying this line to the following:

\medskip
\noindent \texttt{if \textbf{(}(token\_ad /= new\_usb\_addr)\\\indent\textbf{and (endp /= "0001"))} or \\\indent(pid /= not token\_in(16 downto 13) then}
\medskip

\noindent
Now our device can process incoming tokens with address mismatches if they are intended for endpoint 1.
We note that with this modification the platform would inject endpoint 1 traffic on behalf of itself and victim devices connected on the bus in the same speed mode.
We thus further alter the IN endpoint 1 \american{behavioral}{behavioural} logic to only send data for probes with address mismatches, allowing the device to ignore its own traffic, sending NAKs to all of its own probes.
We similarly modify the USB 2.0 device implementation to the same effect. 

\section{Testing USB Hubs}\label{sec:hubs}
In this section we evaluate the hardware setups that are vulnerable to our injection attack. 
We evaluate a total of 29 hubs and find that 14 of them are vulnerable.
We then investigate features that correlate with some of these vulnerabilities.
Finally, we test the impact of bus topology on our injection attack.

\subsection{Testing Methodology}\label{s:methodology}
\begin{figure}[htb]
    \centering
    \includegraphics[width=\linewidth]{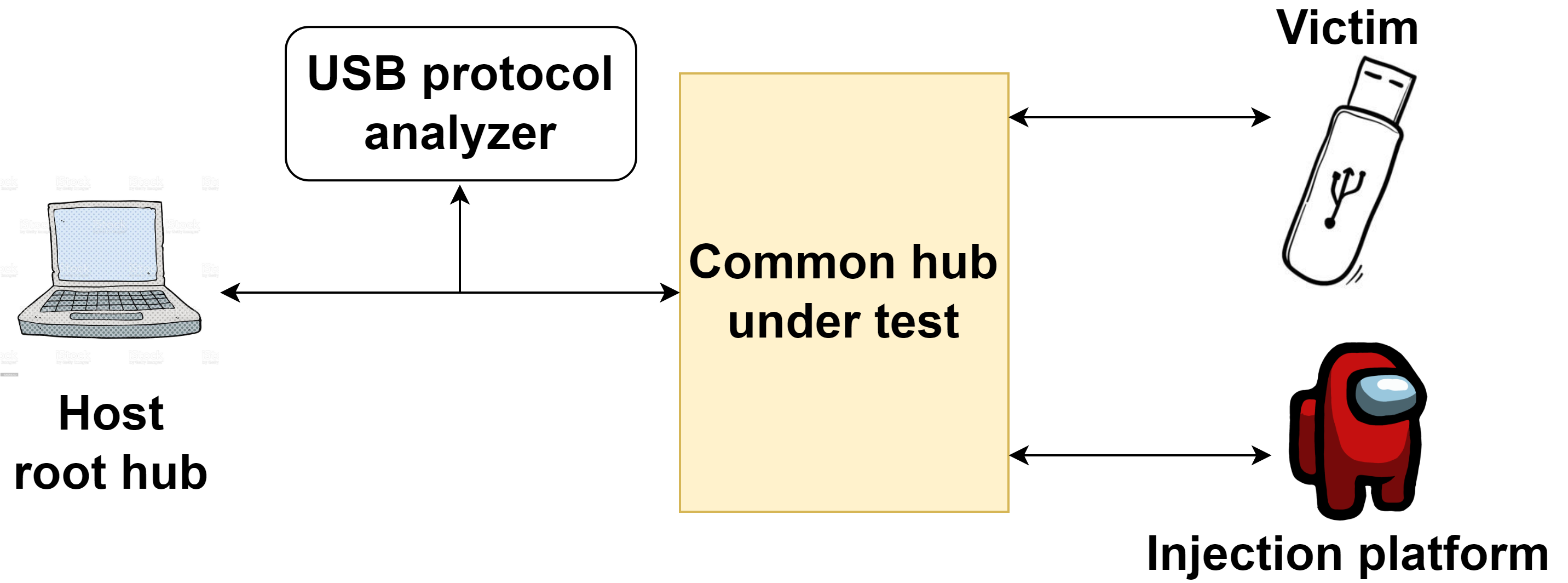}
    \caption{The testing environment.}
    \label{fig:testenv}
\end{figure}

\parhead{Topology.}
To test for injection attacks, we configure our injection platform to inject a unique and easily identifiable data sequence into the USB communication stream.
We then set up an experiment in which both an injection platform and victim device operating at the same speed mode are connected to the host PC through the hub under test. See \cref{fig:testenv}.
We refer to this hub as the \emph{common hub}.

\parhead{Experimental Setup.}
Between the host and the common hub we attach a Totalphase Beagle USB 5000 protocol \american{analyzer}{analyser}, which observes and logs all traffic on the link between the common hub and the host.
Because both the victim device and the injection platform are connected to the host via the common hub, the protocol \american{analyzer}{analyser} also captures all traffic between the host and these devices.
We repeat each experiment three times, once for each of the operating speeds.
For the LS and FS operating speed, we use our USB 1.x injection platform from \cref{s:usb1.x}, with keyboards as the victim devices.
For the HS operating speed we use our USB 2.0 injection platform from \cref{s:usb2.0} with mass storage victim devices.
The specific models are listed in \cref{table:victims} in the appendix.

\parhead{Communication Analysis.}
When our injection platforms inject transmissions during a victim's time slot the genuine victim response is also sent, causing a collision at the common hub.
A vulnerable hub that enables injection continues to forward the first incoming transmission upstream and blocks all subsequently arriving simultaneous transmissions.
In our experiments this result is evident from observing that the protocol  \american{analyzer}{analyser} log attributes the unique data sequence that the injection platform transmits to the victim device's assigned address.
Otherwise, if the hub sends a garble/error sequence upstream when it detects a collision, the unique data sequence does not appear in the protocol \american{analyzer}{analyser}'s log.

\subsection{Targeting USB 2.0 Hubs}\label{s:usb2}
We begin our investigation with a test of 16 USB~2.0 hubs. 
These include thirteen standalone hubs, which are attached via a USB cable to a host, and three embedded hubs that motherboard vendors embedded in their products to increase the number of supported USB ports.
\cref{table:2hubs} in the appendix lists all hub models tested and their advertised IDs. 
Note, the embedded hubs were tested by carrying out the end-to-end attacks as described in \cref{section:attacks}.

\parhead{Injection Results.}
We find that 13 out of the 16 tested hub devices are vulnerable to some form of injection attack. 
One device is vulnerable to both USB 1.x and USB 2.0 injection, seven devices only to USB 1.x injection and four devices only to USB 2.0 injection.
Interestingly, all of the embedded hubs we tested are vulnerable to injection and attacking motherboards that feature them is possible even without any external USB hubs.

\parhead{Anomalous Hub Behavior.}
Three of the hubs we tested exhibit specific \american{behavior}{behaviour} not observed in other hubs.
Two \american{unlabeled}{unlabelled} hubs, marked as USB~2.0 hubs, only operate at USB 1.x speeds.
We were unable to find similar hubs and do not know if the issue affects all hubs of the same models or only the specific ones we tested.
The other hub displaying specific \american{behavior}{behaviour} is the embedded hub in the Micro-Star PRO Z690-A motherboard (marked with (\ding{51}) in \cref{table:2hubs}).
The motherboard has only two exposed ports that seem to show an unbalanced \american{behavior}{behaviour}.
Injection from one of the ports works consistently. However injection from the other is intermittent, with the victim device sometimes winning the race, preventing the injection.
A possible explanation for this unique \american{behavior}{behaviour} is that the hub uses asymmetric downstream port arbitration or switching.

\parhead{DoS Results.}
Finally, we tested the hubs with a DoS attack, in which the injection platform transmits a NAK in response to every probe that the host sends to the victim.
In vulnerable hubs, the injection platform wins the race and the host accepts the injected NAK.
In non-vulnerable hubs, the hub detects the collision and sends an error signal, effectively deleting the victim's response.
Thus, in either case, the denial of service attack is effective.
The only exceptions are the embedded hub described above, where in one configuration the victim sometimes wins the race, overcoming the attack; and multi-TT hubs against 1.x traffic injection, since the downstream probes do not reach the injection platform.

\subsection{Targeting USB 3.x Hubs}
We now turn our attention to USB~3.x hubs.
We tested 13 such hubs, also \american{summarized}{summarised} in \cref{table:2hubs} in the appendix.
For backwards compatibility, USB~3.x hubs consist of two logical hubs, one handling USB~3.x SuperSpeed traffic, while the other handles compatibility with USB 2.0 devices.
As our injection platforms do not operate at SuperSpeed, the experiments only test the internal USB~2.0 hubs.
Interestingly, when enumerated, the internal USB~2.0 hubs identify themselves as USB~2.1 hubs. 
The USB specification~\cite{USB3.0} does not specify a version 2.1.
We suspect that this version number is used to differentiate internal hubs from pure USB~2.0 hubs.

Overall, we find that USB~3.x hubs are less vulnerable to injection attacks, with only one of the 13 we tested allowing injection.
USB~3.x hubs are still vulnerable to denial of service attacks for USB~2.0 victims.
USB~1.x victims can also be attacked with USB~1.x denial of service, but only in the case that the internal hub they connect to is single-TT.

\begin{figure}[h]
    \centering
    \includegraphics[width=0.7\linewidth]{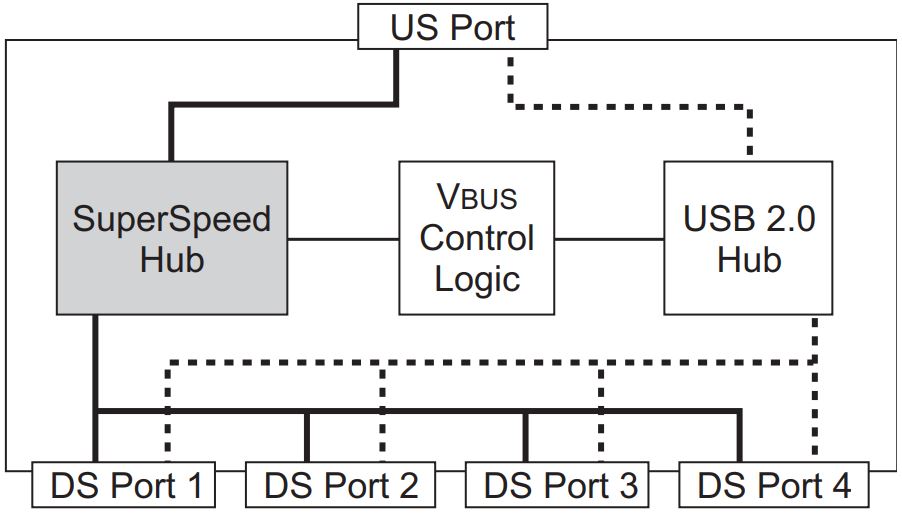}
    \caption{USB 3.0 Hub Architecture~\cite{USB3.0}}
    \label{fig:3hubarch}
\end{figure}

\subsection{Root Hubs}\label{s:roothubs}
Our investigation so far focused on standard hubs that are used for the internal levels of the USB network tree.
\emph{Root hubs} that provide the first tier of attachment points to a host system and are structured differently to standard USB hubs.
Their architecture and operational model is defined in the eXtensible Host Controller Interface for USB (xHCI) specification~\cite{xHCI},  
which \american{standardizes}{standardises} methods for communication between USB software and hardware.
xHCI specifies mechanisms for maintaining port association with connected devices that ultimately routes transmissions to those devices, thus there are no broadcasts of downstream USB 1.x and 2.0 traffic across root hub ports.

As our injection platforms relies on broadcast traffic to time the injection, we cannot expect the attack to work when downstream traffic is not broadcast.
Indeed, we have tested injection using both platforms against multiple xHCI root hubs and found none to be vulnerable because the injectors have no visibility of probes sent to victim devices.

\subsection{The Impact of Transaction Translators}
Observing the results above, we note that the \american{behavior}{behaviour} depends on the configuration of transaction translators in the hub.
Specifically, we find that multi-TT hubs are not vulnerable to USB~1.x injection, because the injection platform does not observe the host's probes.

When performing USB~1.x attacks against single-TT hubs that resist the attack, we observe that the hub sends a `SPLIT-ERR' (error) message upon attempted injection. 
We \american{hypothesize}{hypothesise} that the hub detects the collision and sends the error message instead of a garbled signal.

\subsection{Exploring USB Topologies}\label{s:topologies}
In the testing described so far the victim device and the injection platform have been connected directly to the common hub.
We now investigate the effects that introducing additional tiers of hubs between victim and injection platform has on injection.
USB allows up to 7 hub tiers, where the root hub is the first tier, so up to 5 chained hubs can be cascaded from the root.
In all the experiments, we use a vulnerable hub as the common hub and safe hubs otherwise.

\parhead{USB 2.0 HS Injection.}
We find that is it possible to inject HS traffic in all cases where the injection platform is connected at a hub tier that is not further away from the host than the victim. 
When the injection platform is further than the victim, injection does not work.
We believe the reason to be that the time taken for the HS signal to propagate through a hub repeater is significantly greater than the time by which our injection platform undercuts the victim's response.

\parhead{USB 1.x LS/FS Injection.}
When the common hub is a USB 2.0 or 2.1
hub, we find that injection of USB 1.x traffic only works when both the victim and the injection platform are attached to it directly.
This is because downstream USB~1.x traffic is delivered at USB 2.0 HS (immediately following a SPLIT message) and is only translated to a USB 1.x speed at the hub to which the receiving device is attached.
Consequently, if the victim device and the injection platform (operating as a USB~1.x device) are not connected to the same hub, the injection platform cannot observe broadcasts of translated downstream traffic sent to the victim device.

However, we note that when injecting against a USB~1.x victim, a USB 2.0 injection platform can target the hub the device connects to instead of targeting the device itself.
We verified that, by injecting HS traffic with our USB 2.0 injection platform on behalf of the hub, the injection platform can spoof the hub's translated USB~1.x traffic, thereby confusing the host to accept the injected traffic as if it originated at the USB~1.x victim device.
Therefore, we can inject USB 1.x traffic from the same hub tier only when connected via USB 1.x on the same hub, or we can inject translated HS traffic with a USB 2.0 injection platform connected at a closer hub tier than the victim.
Note that this includes placing the injection platform at the same tier as the hub the victim connects through.

The result is different when using a USB 1.x hub operating at LS/FS speeds as the common hub.
In this case, all hubs connected on the downstream of the common hub revert to operating as USB~1.x hubs.
Consequently, no traffic translation occurs, and the attack works irrespective of the number of intermediate hubs between the common hub and the victim device or the injection platform.

\section{Injection Attacks}
\label{section:attacks}
In previous sections we demonstrated  injection of USB traffic.
We now investigate the security impact of such injection.
Specifically, we demonstrate two attacks, one injecting commands through keystroke data and the other compromising system security by injecting file system contents.

\subsection{Keystroke Command Injection}
\label{section:keyboard}
\parhead{Keyboard USB Stack.}
HID Keyboards typically operate at LS and use endpoint 1 as their main and only input endpoint.
They are simple devices which report character key press and release events.
As such, beyond the USB transaction protocol there is no higher-level protocol used by hosts to elicit data.
Thus, we have directly adapted our USB 1.x injector to demonstrate injection of keystroke commands to the host, like what might be sent in a protocol masquerading attack.

\parhead{Attack Payload.}
In our ad hoc microprocessor application implementation we program a payload of data packets into the platform core directly in hardware and tie their provision to press events for buttons on the board.
The payload sequence opens a Command Prompt on a Windows system.

\parhead{DoS-Switch.}
We further instrument the platform with a `DoS-switch' enabling selective injection of NAKs on behalf of the victim to block its inputs from being forwarded. 
Blocking can be useful under circumstances where the adversary performing injection wants to inject an uninterrupted payload sequence of packets.
The NAKs are sent only when the injector is not providing DATA packets of its own.

\parhead{Experimental Setup.}
We configure our injector to identify as a HID mouse operating at LS.
We connect both the victim and our injection platform to a host through a common hub (one previously found to be vulnerable), again placing the protocol \american{analyzer}{analyser} on the hub's upstream connection.

\parhead{Results.}
We successfully perform keystroke injection attacks against keyboard victims.
We open a Windows Command Prompt, and using the protocol \american{analyzer}{analyser} we observe that the injected traffic is attributed to the victim keyboard's assigned address.
Unplugging the keyboard and pressing the same buttons on our connected injection platform results in no key presses, further confirming that injections have taken place and it has not somehow mistakenly fed keystrokes as a mouse.
We verify the function of our DoS-switch by pressing keys on the victim observing that with the attack enabled, keystrokes do not pass through.

\parhead{Latency.}
Being able to inject successfully means our attack platform can undercut the victim keyboard probe responses.
We confirmed this by inspecting the respective devices' packet timings in the protocol \american{analyzer}{analyser} traffic capture.
We do not expect other keyboards from different manufacturers to have response times fast enough to pose a concern. 
The platform should consistently win transmission races and no further modification is needed to speed up response times.

\parhead{Attacking Gaming Keyboards at FS.}
We alter the USB 1.x injector to work at FS and find that it can also successfully inject against a low latency gaming keyboard at FS with a 1\,kHz poll rate. 
This is because gaming keyboard latency is bounded by the high polling rate and not outright hardware response times. 

\subsection{Hijacking File Transfers}
\label{section:msd}
For our second use case we adapted the USB 2.0 injection platform for compromising the communications of HS \ifFlash flash \else thumb \fi drive victims.
Using it we can hijack device-to-host file transfers.
The platform listens for data requests sent to a \ifFlash flash \else thumb \fi drive victim, which stimulates it to inject data that ultimately alters the contents of files that end up resident on the host.

We further demonstrate this capability in a use case where we compromise a Kali Linux OS image in a boot from USB.

\parhead{Mass Storage Device Stack.}
Here we describe aspects of the mass storage device (MSD) class relevant to the use case.

Endpoint 1 is typically the main data input endpoint for MSDs.
This means we can retain the modifications that made the original HS injection platform.
However, injection has increased complexity in this use case because MSDs implement a \emph{Command/Data/Status} transport protocol across multiple USB transactions and over multiple endpoints.
Its operation is largely analogous to USB's transaction protocol.
This protocol uses the Small Computer System Interface (SCSI)~\cite{SCSI} command format.
For our purposes we are not interested in the information transfers that establish a link between the host and MSD file systems, rather we are only specifically interested in the exchanges that take place during a file transfer on established links.
Once a link has been established, the host periodically sends the SCSI \texttt{Test Unit Ready} (TUR) \emph{Command} which essentially acts as a keep-alive message.
This is followed up by the device sending a \emph{Status} message to the host indicating that it is ready for another command.

\begin{figure}[htbp]
    \centering
    \includegraphics[width=0.65\linewidth]{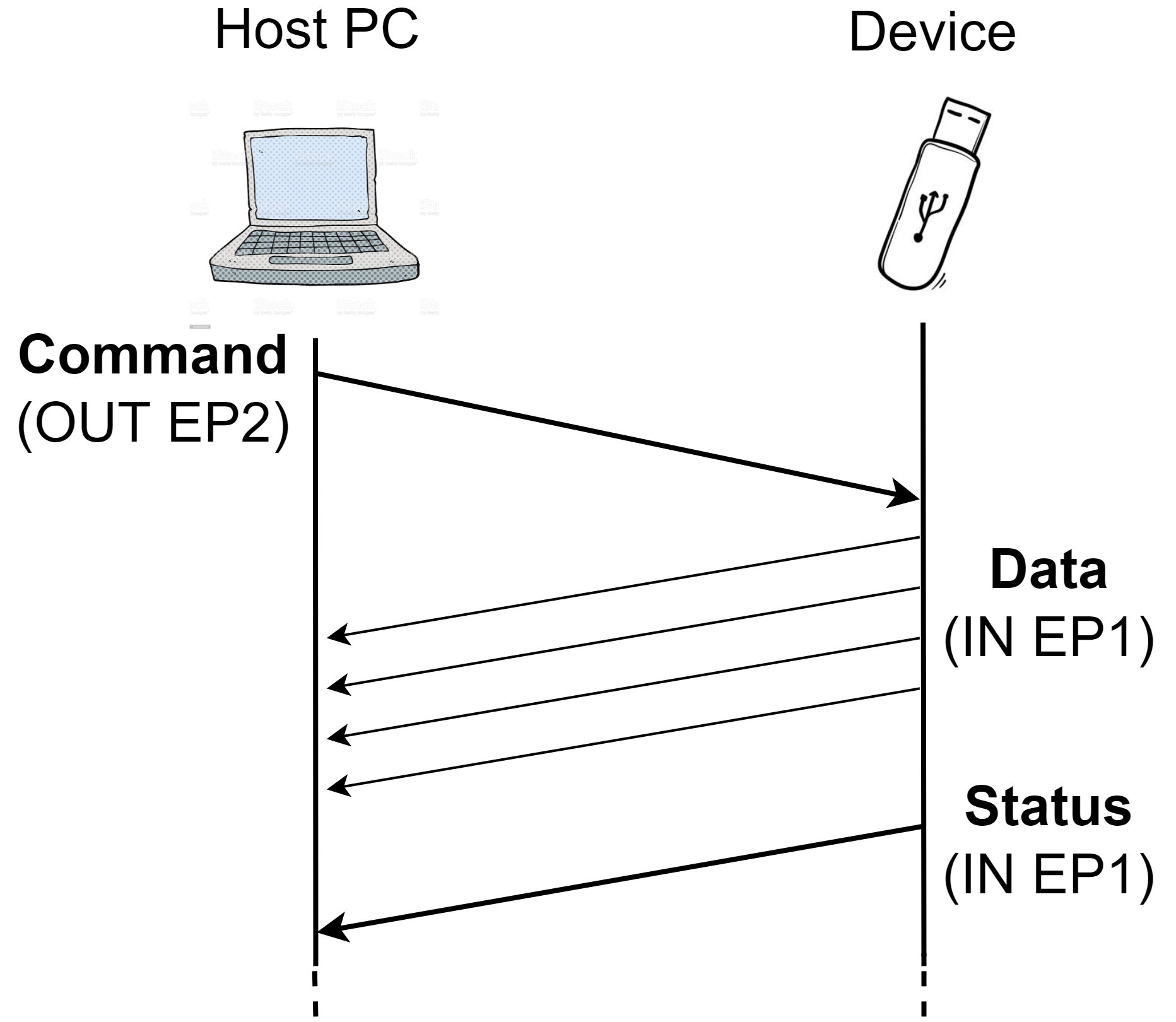}
    \caption{\emph{Command/Data/Status} sequence in a device-to-host mass storage file transfer.
    Each arrow in the diagram is representative of an entire 3-stage USB transaction.}
    \label{fig:commanddatastatus}
\end{figure}

When the host wants to initiate a device-to-host file transfer it will issue a SCSI \texttt{read(10)} \emph{Command} requesting transfer by communicating through OUT endpoint 2, illustrated in \cref{fig:commanddatastatus}.
Among the fields within this \emph{Command} message are the \texttt{read(10)} opcode \texttt{[0x28 0x0a]}, the requested data address offset, the transfer size, and a unique tag.
The file contents are then provided in subsequent \emph{Data} block(s) through one or multiple (depending on size) IN endpoint 1 transactions.
The device indicates transfer completion \emph{Status} through a subsequent IN endpoint 1 transaction.
The \emph{Status} message must include the same unique tag issued in the \emph{Command} message.

At HS, the maximum transfer size for a data block is 512 bytes.
When amounts of data in excess of this are requested it is all transmitted over multiple successive IN transactions.
For brevity, any mentions of IN or OUT communications hereon refer to those over endpoints 1 and 2, respectively.

\parhead{Further Modifications to the Injection Platform.}
Operating with awareness of the extra communication protocol layer of MSD victims requires some higher-level control functionality on the part of our platform.
Directly in the platform hardware we configure ad hoc microcontroller application function to monitor downstream OUT messages to the victim and trigger an internal signal upon detection of the SCSI \texttt{read(10)} command.
We also make it store the transmission size requested and the unique message tag.

We program it to construct injection packets consisting of the \emph{g} character (\texttt{0x67}) (arbitrarily chosen), with the final packet being zero-padded to 512 bytes in length.
It determines how many \emph{g} characters to put in the injected packet(s) from the size field in the triggering \texttt{read(10)} command packet.

Some additional injections after the data is sent are required to fulfil exchanges that complete the file transfer.
We make our platform inject a \emph{Status} transmission (with the registered messaged tag) in response to the IN token following the last \emph{Data} transmission.
We program the platform to then acknowledge the subsequent OUT TUR messages (initial ACK response to the PING token then to OUT message) sent to the victim. 

\parhead{Experimental Setup.}
We configure this injection platform to present itself as a serial communications device operating at HS.
It monitors HS OUT communications to other connected devices and once triggered, injects in response to the subsequent IN tokens addressed to the victim.
The injection platform and victim \ifFlash flash \else thumb \fi drive attach to a common hub (previously found vulnerable) which connects to a Windows host through our protocol \american{analyzer}{analyser}.
We prepare a text file in the victim file system consisting of several different characters.

\parhead{Results.}
The attack is consistently successful since our injector consistently undercuts the MSD responses, and the host terminates the transfer leaving our injected packet contents on the host machine.
We confirm the result with the \american{analyzer}{analyser} capture.
\begin{figure}[htbp]
    \centering
    \includegraphics[width=\linewidth]{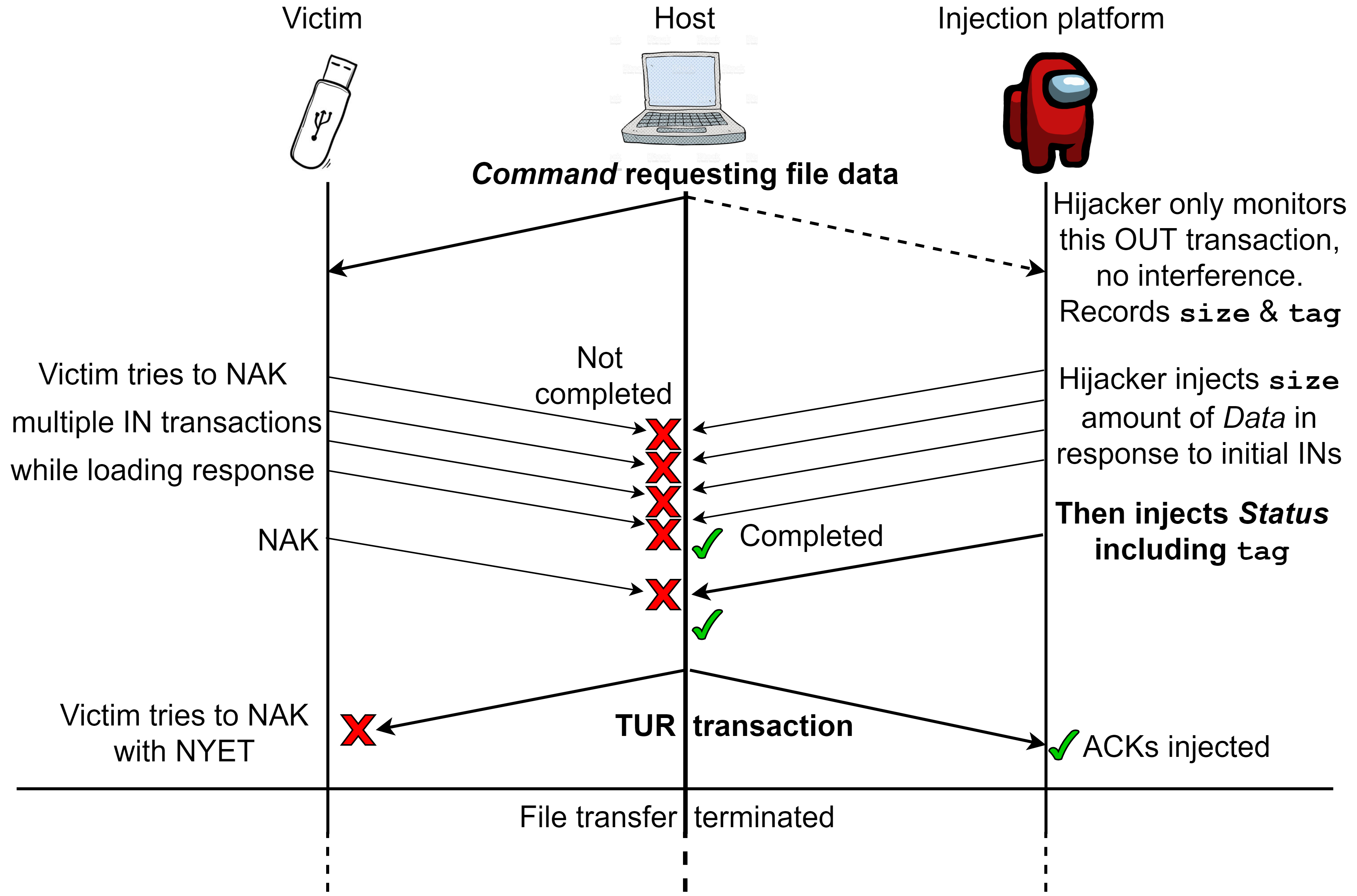}
    \caption{File transfer hijack communications.
    Each complete arrow represents a completed 3-stage USB transaction.
    The dotted line arrow represents a transaction that is merely observed by the injection platform.}
    \label{fig:hijack}
\end{figure}
\cref{fig:hijack} shows bus communications during our file transfer hijack.
After an ACK is injected in response to the TUR command, 
the driver concludes the file transfer.
A key takeaway is that a malicious actor may need to \american{characterize}{characterise} the \american{behavior}{behaviour} of target host system drivers to perform attacks against victims communicating with higher level protocols.

\parhead{Operating System Image Compromise in USB Boot.}
We apply our file transfer hijacker to compromising a Kali Linux OS image transferred when a host boots from USB.
Our attack changes the partition selection option labels that appear in the boot menu, switching the `encrypted persistence' option with the `persistence' option.

To do this, we performed a boot from the victim and recorded all traffic.
In the capture we identify the \texttt{grub.cfg} file contents which happen to be transferred at the 17th IN transaction (and in plaintext) following a request for data at a certain address.
We store the requested address in our platform and alter the injection trigger signal.
The trigger signal is set after our platform has observed 16 IN tokens following a \texttt{read(10)} \emph{Command} requesting the stored address.
We inject a modified version of the genuine packet.

The compromise is successful.
This case is different to the previous file transfer hijack since we are injecting in response to an IN token which is known to be met with a victim DATA packet instead of a NAK.
Our injected data transmission arrives at the host earlier than the victim's and the host later ACKs the victim for the injected transmission, so the victim does not think it needs to resend the data.

Compromise in exactly this manner demands a strong attacker model since it requires knowledge of the victim data transfers that achieve a USB boot, and of course the content transferred.
Nonetheless, it demonstrates that using our injection exploit we can compromise the boot image. 
In a more realistic attack scenario, an injection platform could be configured with the complete capability of a \ifFlash flash \else thumb \fi drive (while still presenting a benign device) and use injection to take over a victim connection entirely to force its own image onto the host.
This is akin to how we previously continued to inject in subsequent communications to terminate a hijacked file transfer.

\section{Circumventing \american{Authorization}{Authorisation} Policies}
\label{section:protections}
Due to the \emph{trust-by-default} nature of USB, attack-capable devices only need access to typical, unprotected computer systems to feed keystroke commands or transfer malicious files.
Attacks of this kind can usually be performed without injections.
However as previously mentioned, injection is useful against protective measures that can govern the \american{authorizations}{authorisations} afforded to connected devices, since these policies are enacted at higher level communication layers than that where injection occurs.
\american{Authorization}{Authorisation} policies trust and process messages that arrive from the host USB controller which, as we have shown, can be wrongly attributed.
Such policies are widely adopted protections according to a survey carried out in 2019~\cite{apricorn} which found over 40\% of \american{organizations}{organisations} to use such policies.

\parhead{Testing USB \american{Authorization}{Authorisation} Policies.}
We test our platforms against various \american{authorization}{authorisation} policies to confirm that they can be circumvented by injection.
Depending on how the policies can be set, for testing we either explicitly allow only a trusted victim to provide input while implicitly blocking all other devices, or we explicitly block our injection platform, allowing all else.
In cases where it is possible to both explicitly allow and block certain devices we do so.
With the policy in operation, we then attempt injection. 
The tested policy solutions and the policies are:

\parhead{\textsc{usbfilter}.} \textsc{usbfilter}~\cite{197251, USBfilter} is a packet-level access control system which can be linked to the Linux kernel.
Through its use, packet filtering rules can be applied to the level of allowing or blocking certain device interfaces, also enabling restriction of interaction between device interfaces with certain applications/processes running on the host.
In testing \textsc{usbfilter}, we allow the trusted devices and block all interaction with our injection platforms.

\parhead{GoodUSB.} GoodUSB~\cite{10.1145/2818000.2818040, GoodUSB} instruments the USB stack to let users moderate which drivers can be loaded for a device based on what functionality they expect, using a popup menu that appears when they plug it in.
We only allow the victims normal use of their expected drivers in testing.
Injection can exploit those victim interfaces both when the attack platforms have been allowed as their other benign device types and when they are not \american{authorized}{authorised} for use.

\parhead{USBGuard.} USBGuard~\cite{USBGuard} is a device access control package. 
This software gives users options to `allow', `block', and `reject' communication with certain devices.
We allow our victims and perform testing where we block and where we reject the injection platforms.

\parhead{Oracle VirtualBox.} the Oracle VM VirtualBox extension pack~\cite{virtualbox} supports the use of USB devices in a virtual machine's guest OS.
The extension lets users maintain a list of devices to be allowed or blocked for use in the VM.
We explicitly allow the trusted device and block the injection platform. 

\parhead{USB-Lock-RP.} USB-Lock-RP~\cite{usblockrp} allows users to selectively allow only certain devices to work on a host.
We allow only our trusted victim \ifFlash flash \else thumb \fi drive/keyboard to operate when testing.
We tested both the free version and a licensed version that Advanced Systems International provided us.

\parhead{Experimental Results.}
We successfully bypass all tested policies.
We believe it is reasonable to claim that injection can be used to bypass all \american{authorization}{authorisation} policies implemented anywhere in the USB software stack.

\parhead{Device Fingerprinting Invariant.}
Device \american{authorization}{authorisation} protections can be bypassed irrespective of the fingerprinting mechanisms they use to identify devices.
Fingerprinting is most commonly based on \emph{VendorID} (VID) and \emph{ProductID} (PID) identifiers~\cite{IDsrepo} supplied by devices during enumeration, however some mechanisms leverage other information like packet timings~\cite{Timeprint} or device electromagnetic emanations~\cite{MAGNETO}. 
As we have demonstrated with selective injection, we can allow victim devices to operate as normal during such fingerprinting processes so that they are correctly identified.

\section{Limitations, Future Work and Countermeasures.}
\parhead{Targeting Additional Devices.}
In its current form, our attack is limited to targeting communications of USB 2.0 and 1.x victim devices.
This is due to our threat model's dependence on monitoring downstream communications, which are only broadcast in these protocol versions, to stimulate injection responses.
However, devices using these protocol versions continue to be highly relevant, as keyboards and other HIDs will continue to be manufactured to the 1.x standard.

The injection platforms created in this work have been made to demonstrate specific use cases attacking keyboard and mass storage device communications.
Our file hijacker for example used specific knowledge of the mass storage device class protocol to achieve desired effects.
We leave the task of implementing attacks against other device types and classes to future work. 

\parhead{Attacking USB 3.x Traffic.}
USB 3.x systems use point-to-point routing which prevents direct off-path communications monitoring, albeit inadvertently since this was introduced as a power saving measure to reduce unnecessary signal transmission in hubs and token address processing at devices.
We do note however, that since USB 3.x hubs incorporate side-by-side SuperSpeed and 2.0 hubs for device backward compatibility~\cite{USB3.0}, the 2.0 (or 2.1) hub within is attacked by our injection exploit, making these 3.x hubs susceptible.

It may yet be possible to inject USB 3.x traffic by transmitting in response to probes that an attacker indirectly monitors, perhaps from signal crosstalk leakages.
Future work can try to target USB 3.x victims with a similar attack model and investigate mechanisms for off-path monitoring of downstream traffic in 3.x hubs.

\parhead{Mitigations.}
A straightforward countermeasure is to use hubs that block transmission when a collision is detected.
This is explicitly mentioned as a design option in the USB 2.0 specification however 
our results show the majority of USB 2.0 hubs do not implement this option.
Conversely, the majority of USB 2.1 hubs (i.e. USB 2.0 hubs within 3.x hubs) do.
Although collision detection prevents our attack, it does not address the core vulnerability as it relies on the trusted device causing a collision.
Nonetheless, collision detection should be effective so long as the trusted device does not malfunction or have a malfunction induced by other means.

Further, adding capability to interpret the garbles sent upstream indicating collision detection in a host's software stack would mean the system user can be alerted that either an injection or malfunction is potentially occurring.
Host controller chips must be designed to pass on collision information to the USB software stack for user alerts to work.

For vulnerable hubs, physical separation of devices from USB port lines through proxies and/or port blocking add-ons can be used to prevent injection and bus sniffing attacks.

\section{Conclusions}
\label{section:conclusion}

In this paper we present an off-path transmission injection attack on the integrity of USB communications, the first of its kind.
This provides us with a complete picture of the threat off-path devices can pose.
We find the majority of USB 2.0 hubs to be vulnerable as they enable injection, and we discovered that a small proportion of USB 3.x hubs are vulnerable.
By using our injection technique, an attacker can circumvent device \american{authorization}{authorisation} policies enforced in a computer's software stack to exploit any communication channels that are trusted according to the user-configured policies.
We circumvent all policies tested.
We further demonstrate two attack scenarios: keystroke command injection and hijacking file transfers. The former allows us to inject malicious commands, and the latter allows us to compromise an OS image when booting from a trusted USB \ifFlash flash \else thumb \fi drive.

While our attacks do require a malicious device connected to the target system, we argue that these attacks do pose a non-trivial threat in some scenarios, especially for high-security USB applications.
As current generation of USB hardware cannot be updated to prevent command injection, we leave the task of designing and deploying USB usage policies for preventing injection attacks to future work.

\ifAnon
\else
\section*{Acknowledgements}
We thank the reviewers from the USENIX Security Program Committee for their insightful feedback, this paper was greatly improved across revisions based on their comments.

This work was supported by
the Air Force Office of Scientific Research (AFOSR) under award number FA9550-20-1-0425; 
an ARC Discovery Early Career Researcher Award number DE200101577; 
an ARC Discovery Project number DP210102670; 
the Blavatnik ICRC at Tel-Aviv University; 
the Defence Science and Technology Group (DSTG), Australia under Agreement ID10620; 
the National Science Foundation under grant CNS-1954712 
and gifts from AMD, Intel, and Qualcomm.
\fi


\appendix

\section*{Appendix}
In this appendix we provide further tables and figures that augment and complete the information in the paper.

\begin{table*}[ht!]
  \footnotesize
  \caption{USB victim devices used to test injection}
\label{table:victims} 
\centering
  \begin{tabular}{lllll}
\toprule
\textbf{Device} & \textbf{\textit{VendorID}} & \textbf{\textit{ProductID}} & \textbf{\textit{bcdDevice}} & \textbf{Speed} \\
\midrule 
Dell Quietkey Keyboard & 0x413C & 0x2106 & 0x0101 & LS \\
Corsair K55 RGB PRO Gaming Keyboard & 0x1B1C & 0x1BA4 & 0x0101 & FS \\
Emtec USB DISK 2.0 & 0x6557 & 0x0121 & 0x0100 & HS \\
Silicon Motion, Inc. USB Flash Disk & 0x090C & 0x1000 & 0x1100 & HS \\
SanDisk U3 Cruzer Micro Flash Drive & 0x0781 & 0x5406 & 0x0200 & HS \\
SanDisk Cruzer Blade Flash Drive & 0x0781 & 0x5567 & 0x0100 & HS \\
\bottomrule
\end{tabular}
\end{table*}

\parhead{Victim Devices Used in Testing.}
\cref{table:victims} lists the devices we used as the victim devices for injection tests.
Overall, we used six victim devices.
For the USB~1.x experiments we used two keyboards, one operating at LS and the other at FS.
For the USB~2.0 experiments we used four different USB disks.

\parhead{Hubs Tested.}
\cref{table:2hubs} lists all hubs we tested.
The model describes the packaging of the hub.
\textbf{Type} indicates whether the hub is a standalone device (S) or embedded within a host system (E), for example embedded within a motherboard.
The ticks in the columns labelled by speed mode (\textbf{1.x} -- LS/FS and \textbf{2.0} -- HS) indicate hubs that were found vulnerable to injection in that speed mode.
\textit{VendorID} (\textbf{\textit{VID}}), \textit{ProductID} (\textbf{\textit{PID}}), and
\textit{bcdDevice} (\textbf{\textit{bcdDev}} -- device version) are descriptors provided by the hub during enumeration with a host.
\textbf{TT} indicates whether the hub is a multi-TT (M) or single-TT (S) system.
Also note that for the 3.0 hubs, we refer to the internal 2.0 hub chip descriptors.

\begin{table*}[ht!]
  \scriptsize
  \caption{USB hub models tested.
  \textbf{Version} indicates USB version.
  \textbf{Type} indicates standalone device (S) or embedded (E) within a host system, e.g. motherboard.
  \textit{VendorID} (\textbf{\textit{VID}}), \textit{ProductID} (\textbf{\textit{PID}}), and \textbf{\textit{bcdDevice}} (device version) are hub-provided descriptors.
  \textbf{TT} indicates single- (S) or multi-TT (M).
  Ticks under \textbf{1.x} or \textbf{2.0} indicates vulnerability to injection in those modes.}
\label{table:2hubs} 
\centering
  \begin{tabular}{lcclllccc}
\toprule
\textbf{USB Hub Model} & \textbf{Version} & \textbf{Type} & \textbf{2.0 Chip Vendor (\textit{VID)}} & \textbf{\textit{PID}} & \textbf{\textit{bcdDev}} & \textbf{TT} & \textbf{1.x} & \textbf{2.0}\\
\midrule 
1PortUSB Network with 3Port & 2.0 & S & Terminus Technology Inc. (0x1A40) & 0x0101 & 0x0100 & S & &\\
D-Link DUB-H7 7-Port & 2.0 & S & Terminus Technology Inc. (0x1A40) & 0x0101 & 0x0111 & M & & \ding{51}\\ 
D-Link DUB-H7 7-Port & 2.0 & S & D-Link Corp. (0x2001) & 0xF103 & 0X0100 & S & \ding{51} & \ding{51} \\
Dell EMC & 2.0 & S & AMECO Technologies (0x214B) & 0x7000 & 0x0100 & S & & \\
Gigabyte B550 AORUS ELITE V2 Motherboard & 2.0 & E & Genesys Logic Inc. (0x05E3) & 0x0608 & 0x8536 & S & \ding{51} & \\
Gigabyte H470 HD3 Motherboard & 2.0 & E & Genesys Logic Inc. (0x05E3) & 0x0608 & 0x8536 & S & \ding{51} & \\
J. Burrows High-Speed & 2.0 & S & Terminus Technology Inc. (0x1A40) & 0x0101 & 0x0111 & M & & \ding{51}\\ 
Micro-Star PRO Z690-A Motherboard & 2.0 & E & Genesys Logic Inc. (0x05E3) & 0x0608 & 0x6070 & S & (\ding{51}) & \\
Speedlink Barras Supreme Hub and Sound Card & 2.0 & S & Genesys Logic Inc. (0x05E3) & 0x0608 & 0x8536 & S & \ding{51} & \\
Startech Industrial 4 Port Mountable & 2.0 & S & NEC Corp. (0x0409) & 0x005A & 0x0100 & S & \ding{51} & \\
Targus & 2.0 & S & Genesys Logic Inc. (0x05E3) & 0x0608 & 0x8537 & S & \ding{51} & \\
Tripp Lite & 2.0 & S & Terminus Technology Inc. (0x1A40) & 0x0101 & 0x0111 & M & & \ding{51}\\
\american{Unlabeled}{Unlabelled} operating at LS/FS & 2.0 & S & Genesys Logic Inc. (0x05E3) & 0x0606 & 0x0702 & S & \ding{51} & N/A \\ 
\american{Unlabeled}{Unlabelled} `Slim' hub operating at LS/FS & 2.0 & S & Genesys Logic Inc. (0x05E3) & 0x0606 & 0x0702 & S & \ding{51} & N/A \\
\american{Unlabeled}{Unlabelled} & 2.0 & S & MOAI Electronics Corp. (0x14CD) & 0x8601 & 0x0000 & S & &\\ 
\american{Unlabeled}{Unlabelled} & 2.0 & S & Terminus Technology Inc. (0x1A40) & 0x0101 & 0x0111 & M & & \ding{51}\\
Alogic USB-C Fusion SWIFT 4-in-1 & 3.0 & S & Genesys Logic Inc. (0x05E3) & 0x0610 & 0x0655 & S & \ding{51} & \\
Asmedia ASM107x & 3.0 & S & QNAP System Inc. (0x1C04) & 0x2074 & 0x0100 & M & & \\
Belkin F4U090 & 3.0 & S & Belkin International, Inc. (0x050D) & 0x090B & 0x5102 & M & &\\
Bonelk 4 Port & 3.0 & S & Realtek Semiconductor Corp. (0x0BDA) & 0x5411 & 0x0101 & M & &\\
Channel+ & 3.0 & S & VIA Labs Inc. (0x2109) & 0x2813 & 0x0221 & S & &\\
HP & 3.0 & S & HP Inc. (0x03F0) & 0x444A & 0x0125 & M & &\\
Plugable & 3.0 & S & VIA Labs Inc. (0x2109) & 0x2813 & 0x9011 & S & &\\
Sabrent & 3.0 & S & Genesys Logic Inc. (0x05E3) & 0x0610 & 0x0655 & S & &\\
Satechi & 3.0 & S & VIA Labs Inc. (0x2109) & 0x2817 & 0x0383 & M & &\\
Smart Sync \& Charge & 3.0 & S & Genesys Logic Inc. (0x05E3) & 0x0610 & 0x9226 & M & &\\
Targus & 3.0 & S & VIA Labs Inc. (0x2109) & 0x2813 & 0x9011 & S & &\\
Ugreen & 3.0 & S & VIA Labs Inc. (0x2109) & 0x2817 & 0x9013 & M & &\\
Wavlink & 3.0 & S & VIA Labs Inc. (0x2109) & 0x2815 & 0x0704 & M & &\\
\bottomrule
\end{tabular}
\end{table*}

\begin{table*}[ht!]
  \footnotesize
  \caption{USB product vendors and manufacturers contacted for disclosure}
\label{table:disclosure} 
\centering
  \begin{tabular}{llcl}
\toprule
\textbf{Company} & \textbf{Type} & \textbf{Report Sent} & \textbf{Most recent response to Report}\\
\midrule 
Alogic & Hub vendor & \ding{51} & Technical team assessing report findings \\
D-Link & Hub vendor and manufacturer & \ding{51} & Asserted their products are specification compliant, \\
 & & & technical team assessing report findings \\
Genesys Logic Inc. & Hub manufacturer & & No response to initial contact \\
Gigabyte Technology & Hub vendor & \ding{51} & Technical team assessing report findings \\
J. Burrows \ifAnon\else(Officeworks)\fi & Hub vendor & & No response to initial contact \\ 
Oracle (VM Virtual Box) & Software vendor & & No response to initial contact \\
NEC Corporation & Hub manufacturer & & No response to initial contact \\
Speedlink & Hub vendor & \ding{51} & Not concerned because the product has been discontinued\\
Startech & Hub vendor & \ding{51} & Technical team assessing report findings \\
Targus & Hub vendor & \ding{51} & Escalated with senior management \\
Terminus Technology Inc. & Hub manufacturer & \ding{51} & Not concerned, they say it is irrelevant with their hubs \\
Tripp Lite & Hub vendor & \ding{51} & Not concerned because their product is not a network device\\
USB-Lock-RP & Software vendor & \ding{51} & Acknowledged findings \\
\bottomrule
\end{tabular}
\end{table*}

In some cases duplicate chips are listed, for instance the 2.0 hub chip within the vulnerable Alogic 3.0 hub has the same IDs as a chip from a non-vulnerable model - the Sabrent 3.0 hub.
Despite the identical IDs we found these two chips to present slightly different descriptor sets which indicates they may have undergone different configuration when ported into their USB 3.0 hub.
For this reason we have retained all cases of duplicate chips among different hub products.

\parhead{Responsible Disclosure.}
\cref{table:disclosure} lists the vendors we notified and summarises their responses.
To ensure that disclosure is coordinated, the initial contact did not include a bug report.
Instead, it sought an agreement not to disclose until a mutually agreed disclosure date.
Of the 13 vendors we approached, four failed to respond to the initial contact, despite repeated attempts. 
The other nine agreed to the terms and received the report.
Only four of the vendors responded to the contents of the report, three of which claim not to be concerned, whereas one vendor acknowledged the finding.

\begin{figure}[h]
    \centering
    \includegraphics[width=\linewidth]{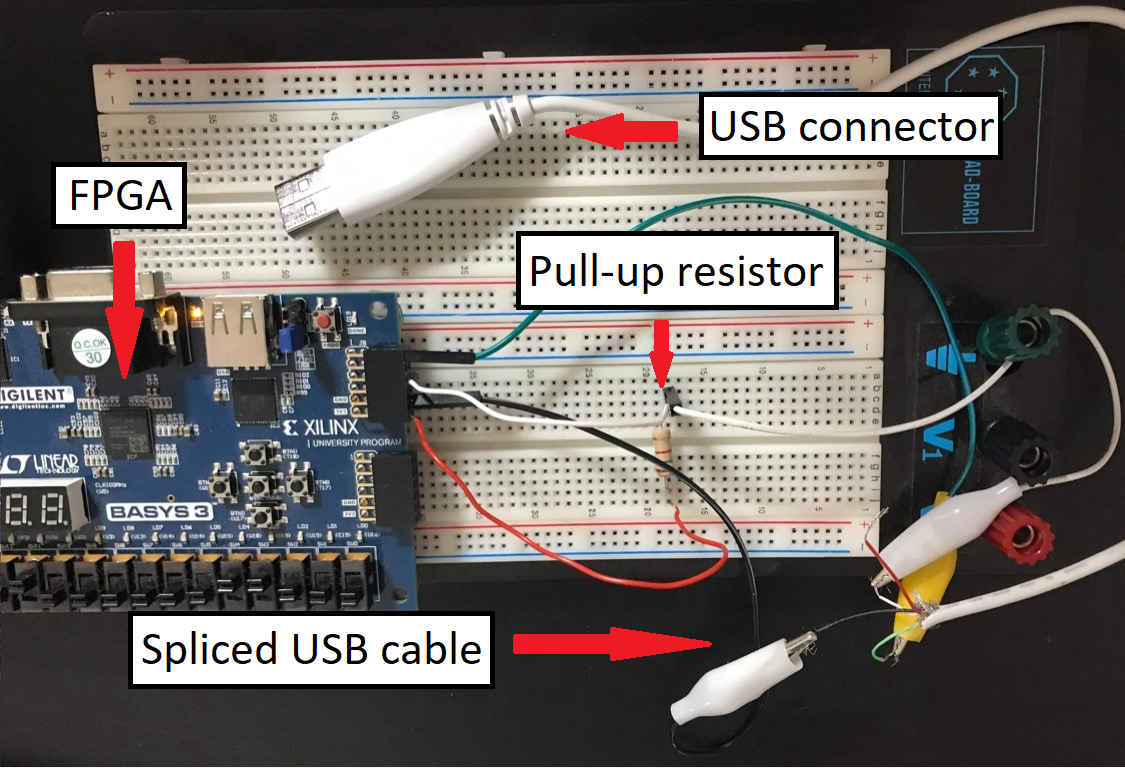}
    \caption{USB 1.x target hardware}
    \label{fig:LSplat}
\end{figure}

\begin{figure}[h]
    \centering
    \includegraphics[width=\linewidth]{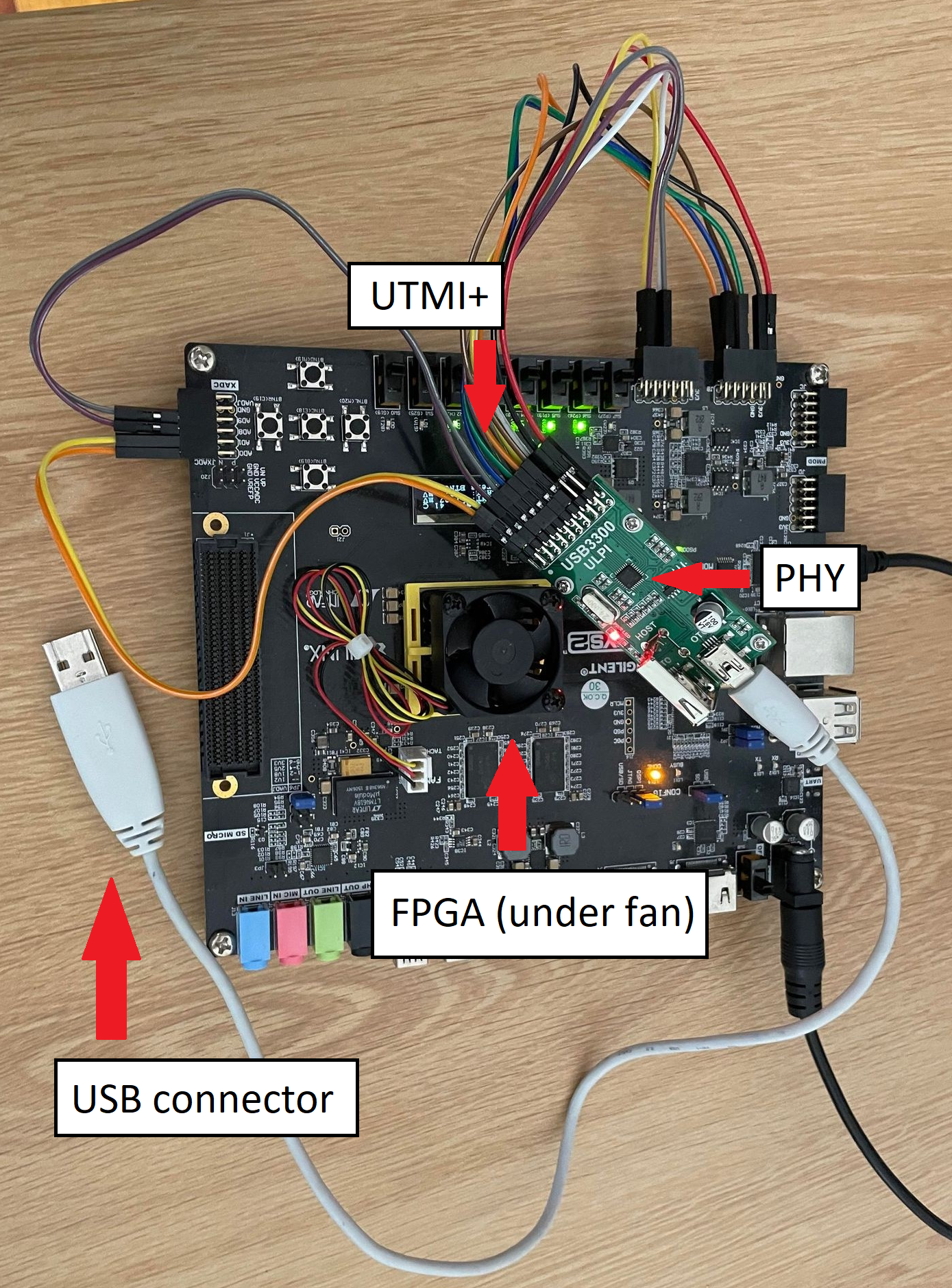}
    \caption{USB 2.0 target hardware}
    \label{fig:HSplat}
\end{figure}

\parhead{Injection Platform Hardware.}
Last, we include images of the target hardware to which we ported our injection platform implementations.
\cref{fig:LSplat} shows the 1.x injector hardware described in \cref{s:usb1.x}.
This includes additional external circuitry required to interface directly to the USB lines.
\cref{fig:HSplat} shows the 2.0 injector hardware described in \cref{s:usb2.0}.
The hardware comprises a target FPGA configured with the USB device core, and an external dedicated PHY module connected by wires over UTMI+.

\end{document}